\documentclass[fleqn,usenatbib]{mnras}

\usepackage{newtxtext,newtxmath}
\usepackage{float}

\usepackage[T1]{fontenc}

\DeclareRobustCommand{\VAN}[3]{#2}
\let\VANthebibliography\thebibliography
\def\thebibliography{\DeclareRobustCommand{\VAN}[3]{##3}\VANthebibliography}

\usepackage{graphicx}	
\usepackage{amsmath}	

\usepackage{ae,aecompl}
\usepackage{relsize}
\usepackage{upgreek}
\usepackage{bm}
\usepackage{ulem}

\graphicspath{{./}{figures/}}

\title[The Dynamical State of Massive Clumps]{The Dynamical State of Massive Clumps}

\author[Z.-J. Lu et al.]{Zu-Jia Lu,$^{1}$\thanks{E-mail:luzujia@icc.ub.edu}
Veli-Matti Pelkonen,$^{1}$
Mika Juvela,$^{2}$
Paolo Padoan,$^{1,3}$
Troels Haugb{\o}lle$^{4}$
and \AA ke Nordlund$^{4}$
\\
$^{1}$Institut de Ci\`{e}ncies del Cosmos, Universitat de Barcelona, IEEC-UB, Mart\'{i} i Franqu\`{e}s 1, E08028 Barcelona, Spain\\
$^{2}$Department of Physics, PO Box 64, University of Helsinki, 00014, Helsinki, Finland\\
$^{3}$ICREA, Pg. Llu\'{i}s Companys 23, 08010 Barcelona, Spain\\
$^{4}$Niels Bohr Institute, University of Copenhagen, {\O}ster Voldgade 5-7, DK-1350 Copenhagen K, Denmark
}

\date{Accepted XXX. Received YYY; in original form ZZZ}

\pubyear{2021}

\begin{document}
\label{firstpage}
\pagerange{\pageref{firstpage}--\pageref{lastpage}}
\maketitle


\begin{abstract}

The dynamical state of massive clumps is key to our understanding of the formation of massive stars. In this work, we study the kinematic properties of massive clumps using synthetic observations. We have previously compiled a very large catalog of synthetic dust-continuum compact sources from our 250 pc, SN-driven, star formation simulation. Here, we compute synthetic $\rm N_{2}H^{+}$ line profiles for a subsample of those sources and compare their properties with the observations and with those of the corresponding three-dimensional (3D) clumps in the simulation. We find that the velocity dispersion of the sources estimated from the $\rm N_{2}H^{+}$ line is a good estimate of that of the 3D clumps, although its correlation with the source size is weaker than the velocity-size correlation of the 3D clumps. The relation between the mass of the 3D clumps, $M_{\rm main}$, and that of the corresponding synthetic sources, $M_{\rm SED}$, has a large scatter and a slope of 0.5, $M_{\rm main} \propto M_{\rm SED}^{0.5}$, due to uncertainties arising from the observational band-merging procedure and from projection effects along the line of sight. As a result, the virial parameters of the 3D clumps are not correlated with the clump masses, even if a negative correlation is found for the compact sources, and the virial parameter of the most massive sources may significantly underestimate that of the associated clumps. 

\end{abstract}


\section{Introduction}

The formation of low-mass stars has been characterized by the evolution of their spectral-energy distribution (SED), classified according to four main classes, from Class 0 to Class III \citep{Myers+87,Lada87,Adams+87,Andre+93}. Although the precise chronology is not straightforward, it is well established that different classes represent a time evolution, from collapsing cores to pre-main-sequence stars. A corresponding observational classification is not available for the early phases of massive stars, whose formation remains shrouded in mystery. The problem is often attributed to the short formation timescale of massive stars.

In the core-collapse picture \citep{McKee+Tan02,McKee+Tan03}, a massive prestellar core collapses into a massive protostar in a time of order $\sim 100$~kyr. Because of its very short Kelvin–Helmholtz contraction timescale, the protostar turns rapidly into a main sequence star, and its strong radiative feedback quickly disperses the parent environment, leaving few clues of the initial conditions. Numerical simulations of star formation in turbulent clouds usually present quite the opposite picture, where massive stars take longer to form than lower-mass stars \citep[e.g.][]{Bonnell+2004,Smith+09,Wang+10,Padoan+14luminosity,Padoan+20massive}. In a recent work, \citet{Padoan+20massive} have proposed a scenario referred to as the {\it inertial-inflow model}, built around this numerical evidence. In this scenario, as in the numerical simulations, massive stars form from the gradual accretion of gas by intermediate-mass stars, over a relatively long timescale, as prestellar cores tend to have a characteristic mass significantly smaller than that of massive stars \citep[see also][for a related scenario based on the observations]{Tige+17,Motte+18}.

Irrespective of the specific formation scenario, massive stars can only be formed in the presence of a large mass reservoir. Massive clumps have played an important role in the study of the origin of massive stars, thanks to the selection of large samples associated with ultracompact HII regions, water masers, or other tracers of the early feedback from massive stars \citep[e.g.][]{Cesaroni+88,Wood+Churchwell89,Wood+Churchwell89b,Sridharan+02,Beuther+02,Shirley+03}. At the spatial resolution of single-dish surveys, active massive clumps usually host the formation of stellar clusters, rather than of a single star, which complicates the interpretation of the observations. At the higher resolution of recent ALMA surveys, it is apparent that massive clumps are highly fragmented, as expected from their supersonic line-widths, and it becomes possible to identify individual prestellar cores \citep[e.g.][]{Sanhueza+17,Contreras+18,Motte+18,Sanhueza+19,Li+19,Servajean+19,Kong19cmf,Pillai+19,Svoboda+19}. While these recent studies highlight new problems, such as the absence of high-mass prestellar cores, the statistics of large samples of massive clumps from single-dish surveys are still key to understand the origin of massive stars through the comparison with high-dynamic-range simulations.   

The Millimetre Astronomy Legacy Team 90 GHz Survey (MALT90) \citep{Jackson+2013,Contreras+17} and the Herschel Infrared Galactic Plane Survey (Hi-GAL) \citep{Molinari+2016,Elia+2017,Elia+21} have produced the largest catalogs to date of massive clumps. Follow-up studies combining the two surveys have further characterized the dynamics of some of the clumps, such as their line-widths and infall rates \citep[e.g.][]{Traficante+17,Traficante+18}. In \citet{Lu+2021}, we compiled a very large synthetic catalog of compact sources, using a star-formation simulation of a 250~pc region of the ISM driven by supernova (SN) explosions \citep{Padoan+SN1+2016ApJ,Padoan+SN3+2016ApJ,Padoan+17_SN4,Lu+20SN}, and the same source-extraction and photometric procedures as for the Hi-GAL catalog. We showed that our catalog matches the properties of the Hi-GAL sources from the Outer Galaxy. In addition, the comparison with the three-dimensional (3D) datacubes showed that the mass of the compact sources greatly overestimates that of the corresponding 3D clumps. Due to the much higher depth towards the Inner Galaxy, projections effects are exacerbated there. 

In this work, we analyze synthetic $\rm N_{2}H^{+}$ emission line spectra from a subset of protostellar synthetic sources to derive velocity dispersion and virial parameter values and compare them with those from the follow-up studies by \citet[][]{Traficante+17,Traficante+18}. Our main goal is to assess how the mass discrepancies and projection effects found in \citet{Lu+2021} over the whole catalog impact the determination of the dynamical state of this subset of protostellar clumps. 
Throughout the paper, the term {\it (compact) source} refers to an object identified from synthetic or real observations, whose properties are measured from dust continuum or molecular emission line maps, while {\it (3D) clump} refers to the actual three-dimensional clump corresponding to a compact source, whose properties can be measured directly only in simulation datacubes.   
The structure of the paper is as follows. In \S~\ref{simulation}, we summarize the numerical simulation. Radiative transfer, synthetic observations, the criteria for the source selection and the derivation of the velocity dispersion are described in \S~\ref{synthetic}. In \S~\ref{comparison_obs} and \S~\ref{comparison_mhd}, we compare the results from the synthetic observations with those from the Hi-GAL sources and from the corresponding 3D clumps. Our findings are discussed in \S~\ref{discussion} and the main conclusions are summarized in \S~\ref{conclusions}.

\section{Simulation} \label{simulation}

This work is based on the same large-scale MHD simulation of star formation used in \citet{Padoan+17sfr} to study the star-formation rate in molecular clouds, in \citet{Padoan+20massive} to study the formation of massive stars, and in \citet{Lu+20SN} to study the effect of SNe on the dispersion of MCs. The simulation has been continuously run, during the past two years, under a multi-year PRACE project, and will be run for another year, until it reaches approximately 100~Myr of evolution. It describes an ISM region of size $L_{\rm box}=250$ pc and total mass $M_{\rm box}=1.9\times 10^6$ $\rm M_{\odot}$, where the turbulence is driven by SNe alone. The 3D MHD equations are solved with the AMR code RAMSES \citep{Teyssier+2002A&A,Fromang+06,Teyssier07}, using periodic boundary conditions. We refer the reader to the papers cited above for details about the numerical setup. In the following, we briefly summarize only the main features relevant to this work.

The energy equation includes the pressure-volume work, the thermal energy introduced to model SN explosions, a uniform photoelectric heating as in \citet{Wolfire+95}, with efficiency $\epsilon=0.05$ and the FUV radiation field of \citet{Habing68} with coefficient $G_0=0.6$ (the UV shielding in MCs is approximated by tapering off the photoelectric heating exponentially above a number density of 200 cm$^{-3}$), and a tabulated optically thin cooling function constructed from the compilation by \citet{Gnedin+Hollon12} that includes all relevant atomic transitions. Molecular cooling is not included, due to the computational cost of solving the radiative transfer. The thermal balance between molecular cooling and cosmic-ray heating in dense gas is emulated by setting a limit of 10~K as the lowest temperature of dense gas. However, to generate synthetic observations of the dust emission, the radiative transfer is computed by postprocessing individual snapshots, including all stars with mass $>2 \, \rm M_{\odot}$ as point sources (see \S~\ref{synthetic}).     

The initial conditions of the simulation are zero velocity, uniform density, $n_{\rm H,0}=5$ cm$^{-3}$, uniform temperature, $T_0=10^4$ K, and uniform magnetic field, $B_0=4.6$ $\mu$G. During the first 45~Myr, self-gravity was not included and SN explosions were randomly distributed in space and time, at a rate of 6.25 SNe Myr$^{-1}$. The resolution was $dx=0.24$ pc, achieved with a $128^3$ root grid and three AMR levels. The minimum cell size was then decreased to $dx=0.03$ pc, using a root-grid of $512^3$ cells and four AMR levels, for an additional period of 10.5 Myr, still without self-gravity. At $t=55.5$ Myr, gravity is introduced and the minimum cell size is further reduced to $dx=0.0076$ pc by adding two more AMR levels. This resolution allows us to resolve the formation of individual massive stars, so the time and location of SNe are computed self-consistently from the evolution of the massive stars. 

Individual stars are modeled with accreting sink particles, created when the gas density is larger than $10^6$ cm$^{-3}$ and other conditions are satisfied \citep[see][for details of the sink particle model]{Haugboelle+2018}. A SN is created when a sink particle of mass larger than 7.5~$\rm M_{\odot}$ has an age equal to the corresponding stellar lifetime for that mass \citep{Schaller+92}. The sink particle is removed and the stellar mass, momentum, and $10^{51}$ erg of thermal energy are added to the grid with a Gaussian profile \citep[see][for further details]{Padoan+SN1+2016ApJ}. By the last simulation snapshot used in this work, corresponding to a time of $34.2$~Myr from the inclusion of self-gravity and star formation, $4,283$ stars with mass $> 2\, \rm M_{\odot}$ have been generated, of which $\sim 415$ have already exploded as SNe. The stellar mass distribution is consistent with Salpeter's IMF \citep{Salpeter55} above $\sim 8\,\rm M_{\odot}$, but is incomplete at lower masses (it starts to flatten at a few solar masses instead of at a fraction of a solar mass), as expected for the spatial resolution of the simulation.

\section{Synthetic Observations} \label{synthetic}

The sources in \citet{Traficante+18} were first selected from Herschel's Hi-GAL compact source catalog \citep{Elia+2017}, taking dust continuum sources which had reliable mass and distance estimates. For that subsample, they then used $\rm N_{2}H^{+}(1-0)$ data from the MALT90 survey of 3-mm emission lines \citep{Jackson+2013} to find sources with high signal-to-noise and with clean spectra (i.e., consistent with a single source in the line of sight, without other sources along the line of sight adding more emission lines). In order to derive a comparison synthetic dataset, we use a synthetic compact source catalog from \citet{Lu+2021}, and synthetic molecular line observations.

\subsection{Synthetic Dust Continuum Maps} \label{synthetic_maps}

We have computed synthetic dust continuum maps using the continuum radiative transfer program SOC \citep{Juvela_2019} in three snapshots of our simulation at times $15.4$, $23.3$, and $34.2 \,\rm Myr$ from the beginning of self-gravity and star formation. These snapshots sample different conditions in the star-formation history of the simulation. The details of the radiative transfer modelling and the creation of the synthetic catalog based on the synthetic dust continuum maps are in \citet{Lu+2021}, and are summarized only briefly here.

The inputs for the SOC program were the density field from the MHD run, the dust model of \citep{OH_1994} that is more appropriate for dense medium, the external radiation field corresponding to the local interstellar radiation field \citep{Mathis_1983}, and point sources in each snapshot corresponding to stars with mass $> 2\, \rm M_{\odot}$ (909, 2431 and 3868 stars, respectively, formed self-consistently in the simulation). SOC was used to calculate the equilibrium dust temperature for each model cell and, based on that information, the surface brightness maps at Herschel's bands (70, 160, 250, 350, and 500~$\mu$m).

The surface brightness maps were resampled to the same pixel sizes as the Hi-GAL maps \citep{Elia+21}: 3.2, 4.5, 6.0, 8.0, and 11.5\,arcsec, for the five bands in the order of increasing wavelength. At each wavelength, the full width at half maximum (FWHM) values of the adopted Gaussian telescope beams were three pixels. We added observational noise to the maps so that, after beam convolution appropriate for the assumed distances (2, 4, 8, and 12\,kpc), the noise was consistent with actual observations \citep[see][for details]{Lu+2021}. The Monte Carlo noise due to the radiative transfer method was a few times below the observational noise. For each of the three snapshots, the surface brightness maps were computed in the three different orthogonal directions and at four different assumed distances, resulting in 36 maps at each wavelength.

\subsection{Synthetic Catalog and Initial Source Selection} \label{sample}

\citet{Lu+2021} created a synthetic compact source catalog using the above surface brightness maps. The sources were extracted with the CuTEx code \citep{Molinari+11}, with the same exact method used to generate the the Hi-GAL compact source catalog \citep{Elia+2017,Elia+21}. First, CuTEx extracted the compact sources (location, size, orientation) and their photometry from each individual surface brightness map. Single-band catalogs of each of 36 snapshot-distance-direction combinations were then merged in wavelength based on whether a shorter wavelength source could be found within the longer wavelength source's footprint. Sources that were successfully merged at least in three neighboring Herschel wavelengths from 160 to 500~$\mu$m were considered good detections and kept in the multi-wavelength catalogs, and sources that had a 70~$\mu$m detection as well were classified as protostellar. Temperatures and masses, $M_{\rm SED}$, were derived via SED fitting, using wavelengths from 160 to 500~$\mu$m. The radius of a source, $R$, is half of the circularized CuTEx FWHM at 250~$\mu$m, which was first deconvolved with the Herschel beam if the FWHM was larger than the beam, or kept unchanged if smaller. Finally, the multi-wavelength catalogs were merged into a single master catalog containing 51,831 sources, of which 44,090 were starless and 7,741 were protostellar. This synthetic catalog was shown to be consistent with the Outer Galaxy portion of the Hi-GAL catalog \citep{Lu+2021}.

The sources studied in this paper are selected from the above synthetic catalog of compact sources, to mimic the selection of \citet{Traficante+18} sources from the Hi-GAL Inner Galaxy catalog \citep{Elia+2017}. As most of the selected sources in \citet{Traficante+18} are protostellar, we first limit our selection only to the protostellar sources. Then, in order to ensure that the selected sources are sufficiently bright in $\rm N_{2}H^{+}(1-0)$ emission, we select sources that have high mean densities of their corresponding 3D clumps (see \S~\ref{comparison_mhd} for the 3D clump definition). To take into account the change of resolution with distance, and hence size and volume, we set the mean density lower limit to depend on the distance: $4 \times 10^4$, $1 \times 10^4$, $2.5 \times 10^3$, and $1.25 \times 10^3 \, \rm cm^{-3}$, for 2, 4, 8, and 12\,kpc, respectively.
Because our synthetic catalog does not have a distance uncertainty, we do not need criteria based on the mass and the distance uncertainties as in \citet{Traficante+18}. 

With the above two criteria, our initial sample contains 1,218 synthetic sources with radii, masses and temperatures from the synthetic catalog.

\subsection{Synthetic Line Spectra Maps} \label{synthetic_lines}

The excitation and the resulting observed spectra of ${\rm N_{2}H^{+}}$ were
calculated with the radiative transfer program LOC \citep{Juvela2020},
using molecular data from the LAMDA database \citep{Schoier2005}. The calculations solved the radiation field within the model volume (including all non-local radiative interactions) and, based on these, the non-LTE excitation between rotational levels. The hyperfine structure of the $J=1-0$ transition was taken
into account by assuming LTE conditions between the hyperfine components
\citep{Keto1990}.

The kinetic temperatures were set equal to the dust temperatures that 
were solved above. This is physically justified at densities at and above 
$\sim 10^5$\,cm$^{-3}$, where the gas-dust collisions bring the gas
temperature within a few degrees of the dust temperature
\citep{Goldsmith2001, Juvela2011}. In our targets, most of the 
${\rm N_{2}H^{+}}$ emission does originate from gas above this density. The
temperature approximation is also more accurate for cold gas, where the
critical density of the ${\rm N_{2}H^{+}}$ $J=1-0$ line approaches the $n({\rm
H}_2)=\sim 10^5$\,cm$^{-3}$ limit. However, there is a range of densities
around $n({\rm H}_2) \sim 10^4$\,cm$^{-3}$ where ${\rm N_{2}H^{+}}$ could have
significant abundance, could be collisionally excited (i.e. sensitive to 
the kinetic temperature), and where the dust temperature will 
underestimate the kinetic temperature. This uncertainty is
probably small compared to, for example, the uncertainty of ${\rm N_{2}H^{+}}$ 
abundances.

\begin{figure*}
\centering
\includegraphics[width=0.49\textwidth]{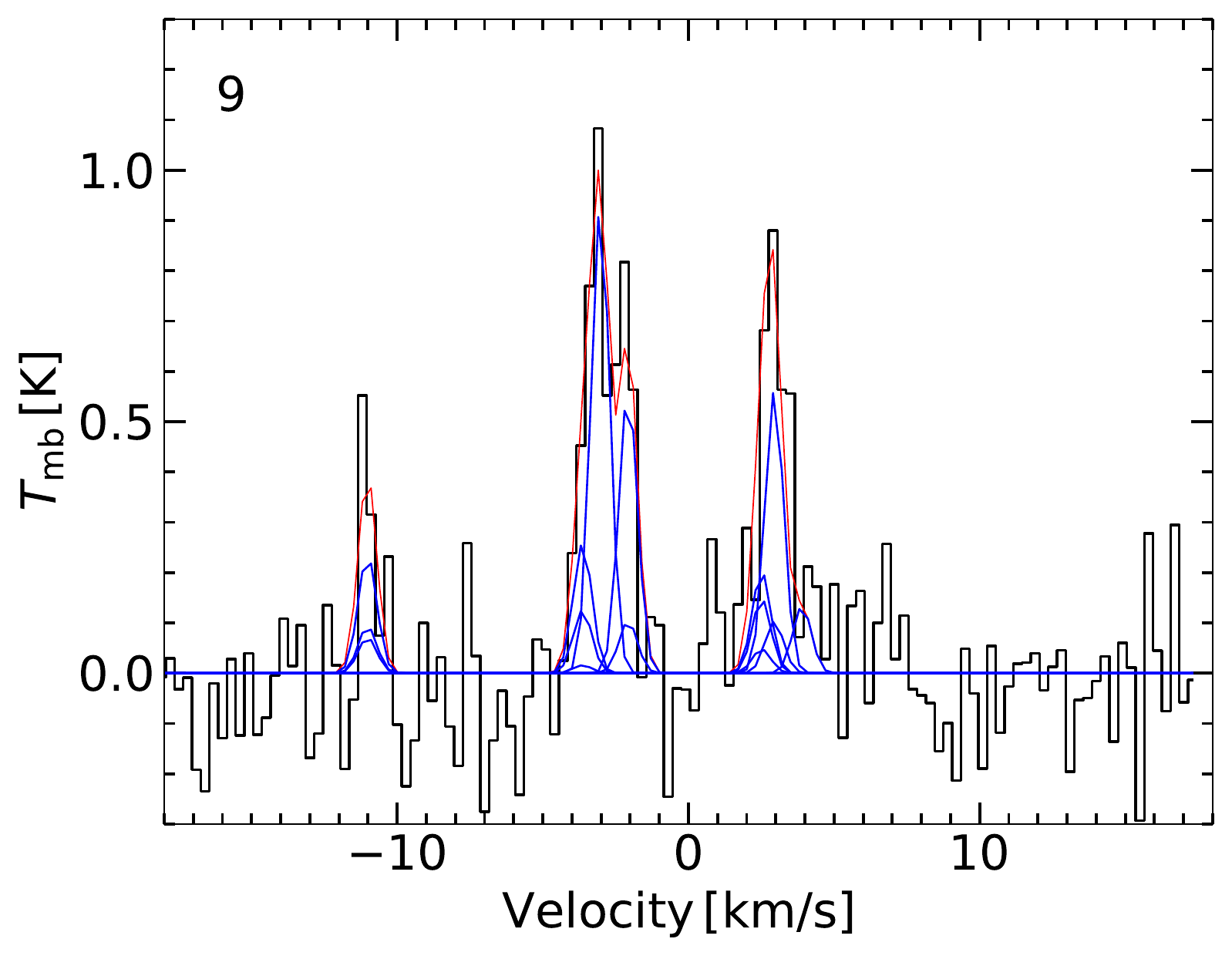}
\includegraphics[width=0.49\textwidth]{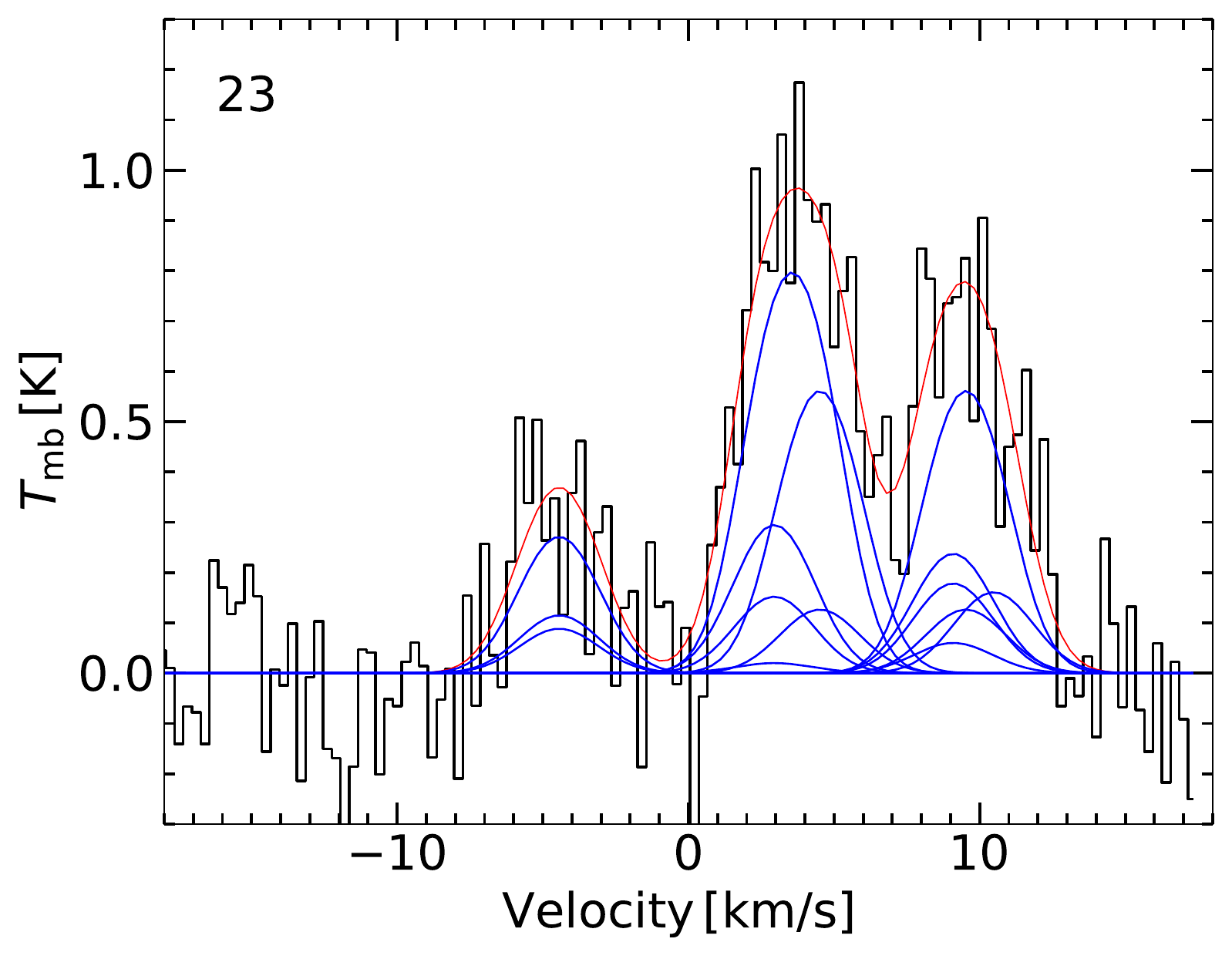}
\caption{$\rm N_{2}H^{+}(1-0)$ line profiles of two compact sources from our synthetic catalog with low (left) and high (right) velocity dispersion. The black line is the synthetic spectrum, and the red line the best fit from PySpecKit. The blue lines are the individual fits to the 15 hyperfine Gaussian components of the $\rm N_{2}H^{+}(1-0)$ line. }
\label{fig_n2hp_sprectra}
\end{figure*}

Estimates of the fractional abundance relative to H$_2$, [${\rm N_2 H^+}$], in star-forming clouds vary
widely, especially in connection with high-mass star formation.
\citet{Ryabukhina2021} found in the infrared dark cloud (IRDC)
G351.78-0.54 values [${\rm N_{2}H^{+}}$]=0.5-2.5$\times10^{-10}$, which are
similar to the results 1-4$\times 10^{-10}$ reported for low-mass cores
\citep{Caselli2002}. However, there are also much larger values. In
\citet{Gerner2014}, the values tended to be slightly below
[${\rm N_{2}H^{+}}$]=$10^{-9}$ for IRDCs and slightly above for high-mass
protostellar objects. In the analysis of data from the MALT90 survey
\citep{Foster2011}, \citet{Miettinen2014} found an average of
[${\rm N_{2}H^{+}}$]=$1.6\times 10^{-9}$, both infrared-dark and infrared-bright sources
being close to this value. For our simulations, we chose the value of
[${\rm N_{2}H^{+}}$]=$1\times 10^{-9}$. The abundances are further scaled with a function $n({\rm H}_2)^{2.45}/(3.0\times 10^8+n({\rm H}_2)^{2.45})$, which
decreases the abundance at densities below $n({\rm
H}_2)=10^{4}$\,cm$^{-3}$. The abundance is thus at $n({\rm
H}_2)=10^{3}$\,cm$^{-3}$ reduced to $\sim 10^{-10}$ and becomes
insignificant in regions of atomic gas. 
The functional form approximates the average density dependence of CO abundances
in the numerical simulations of \citet{Glover2012} \citep[see also][]{Padoan+SN3+2016ApJ}. ${\rm N_2
H^+}$ abundances are likely to follow CO only very approximately, but the initial increase can still
take place at a similar $A_{\rm V}$ or density threshold \citep{Bergin2002}. 
Our results are not sensitive to the exact shape of this function, because the critical density of the ${\rm N_{2}H^{+}}$ $J=1-0$ transition is a few times above the 10$^4$\,cm$^{-3}$ limit and even the effective excitation densities calculated by \citet{Shirley2015} (taking into account photon trapping) are, at the relevant temperatures, within a factor of two.

The radiative transfer calculations were iterated (estimation of the
radiation field and the recomputation of level populations), until the the
changes in the spectra were at a level of 0.1\% per iteration. LOC uses
accelerated lambda iterations, which further ensures that the results
have converged to a high precision.

Synthetic molecular line cubes are created at the same distances (2, 4, 8, and 12\,kpc) as the dust continuum surface brightness maps, smoothing with the beam of the MALT90 survey, 38\,arcsec, to mimic the spatial resolution of the MALT90 observations. The spectra has been calculated with 0.11\,km/s velocity resolution, the same as the MALT90 survey.

\subsection{$\rm N_{2}H^{+}(1-0)$ Spectra and Final Source Selection} \label{synthetic_final}

For the initial sample of 1,218 sources (see \S~\ref{sample}), we derive $\rm N_{2}H^{+}(1-0)$ spectra from the molecular line maps in \S~\ref{synthetic_lines}. The synthetic observation $\rm N_{2}H^{+}(1-0)$ spectra are given in antenna temperature $T_{\rm A}^{\ast}$, and for a synthetic telescope, $T_{\rm mb} = T_{\rm A}^{\ast}$. We average the spectra within an aperture of the MALT90 beam, 38\,arcsec, around the position of the 250\,$\mu$m synthetic source, similarly to \citet{Traficante+18}. We then add Gaussian noise with the amplitude of the typical noise of MALT90 survey, 0.25\,K per 0.11\,km/s velocity channel \citep{Jackson+2013}, before averaging the spectra to a velocity resolution of 0.3\,km/s like \citet{Traficante+18} did. We distinguish the $\rm N_{2}H^{+}(1-0)$ spectra as single component and multiple components, if the spectra show the presence of another molecular line source. Similarly to \citet{Traficante+18}, we exclude the sources whose spectra are contaminated by multiple components.

The hyperfine transitions of the $\rm N_{2}H^{+}(1-0)$ spectra were fit using \texttt{PySpecKit}, a spectroscopic analysis toolkit for astronomy \citep{Ginsburg+2011}, which provided the estimated excitation temperature, optical depth, centroid velocity and velocity dispersion of the Gaussian fit for the line profile of each source. An example of the \texttt{PySpecKit} fit for two of the sources in shown in Figure~\ref{fig_n2hp_sprectra}. The black and red lines represent the original spectrum and Gaussian model fitting, respectively, while the blue lines are components for the all the single Gaussian components of individual transitions. The spectra on the left shows narrow features, and thus a small velocity dispersion, while the one on the right shows wider features, starting to blend the transitions together. In the following, we refer to the velocity dispersion of this Gaussian fit as $\sigma_{\rm N_{2}H^{+}}$. 

\texttt{PySpecKit} also returns an error for the velocity dispersion based on the fit. The errors tend to be small, with a median value of 0.08~km/s. In order to test the validity of this error estimate we have performed the spectral fitting with 50 different noise realizations for each source, and have computed the standard deviation of the 50 values of the velocity dispersion. This Monte-Carlo estimation of the error due to different noise realizations gives a median value of 0.1~km/s, comparable to the median value of the formal \texttt{PySpecKit} error, and is also correlated with the \texttt{PySpecKit} error.

As in \citet{Traficante+18}, we finally exclude sources with a signal-to-noise lower than 5. This results in 145 sources being selected for the further analysis and comparison. The 145 source spectra are shown in Appendix \ref{app_n2hp_sprectra}. The properties of our source sample from both the dust continuum fitting (radius, mass and temperature) and line emission fitting (velocity dispersion) are listed in Table~\ref{app_table_clumps} in Appendix~\ref{app_clump_properties}.

\section{Comparison with the Observations} \label{comparison_obs}

In this section, we compare relations from our synthetic sources selected in \S~\ref{synthetic_final} with those from the sample of Hi-GAL sources in \citet{Traficante+18}, where we have removed the HII regions, as we do not have stellar feedback in our simulation, the starless sources, because we consider only protostellar sources from our synthetic catalog, and any sources closer than 1.5~kpc, as our minimum distance in the synthetic catalog is 2~kpc. The same distance selection was applied when comparing our full synthetic catalog with the full Hi-GAL catalog in \citet{Lu+2021}. Finally, we discard sources with temperatures of 40~K, as that value is the maximum of the fitting range rather than a good fit. After removing those sources, we are left with 135 sources from \citet{Traficante+18} that are compared to our sample of 145 synthetic sources.

As shown in \citet{Lu+2021}, sources from the Inner Galaxy have on average larger surface density than sources from the Outer Galaxy and from our synthetic catalog. Because the sample in \citet{Traficante+18} is made of primarily massive sources from the Inner Galaxy, the comparison in this work reflects that systematic surface density mismatch. However, there is enough overlap in mass and surface density between the two samples to justify this comparison.

\subsection{Mass-Size Relation}

\begin{figure}
\centering
\includegraphics[width=0.48\textwidth]{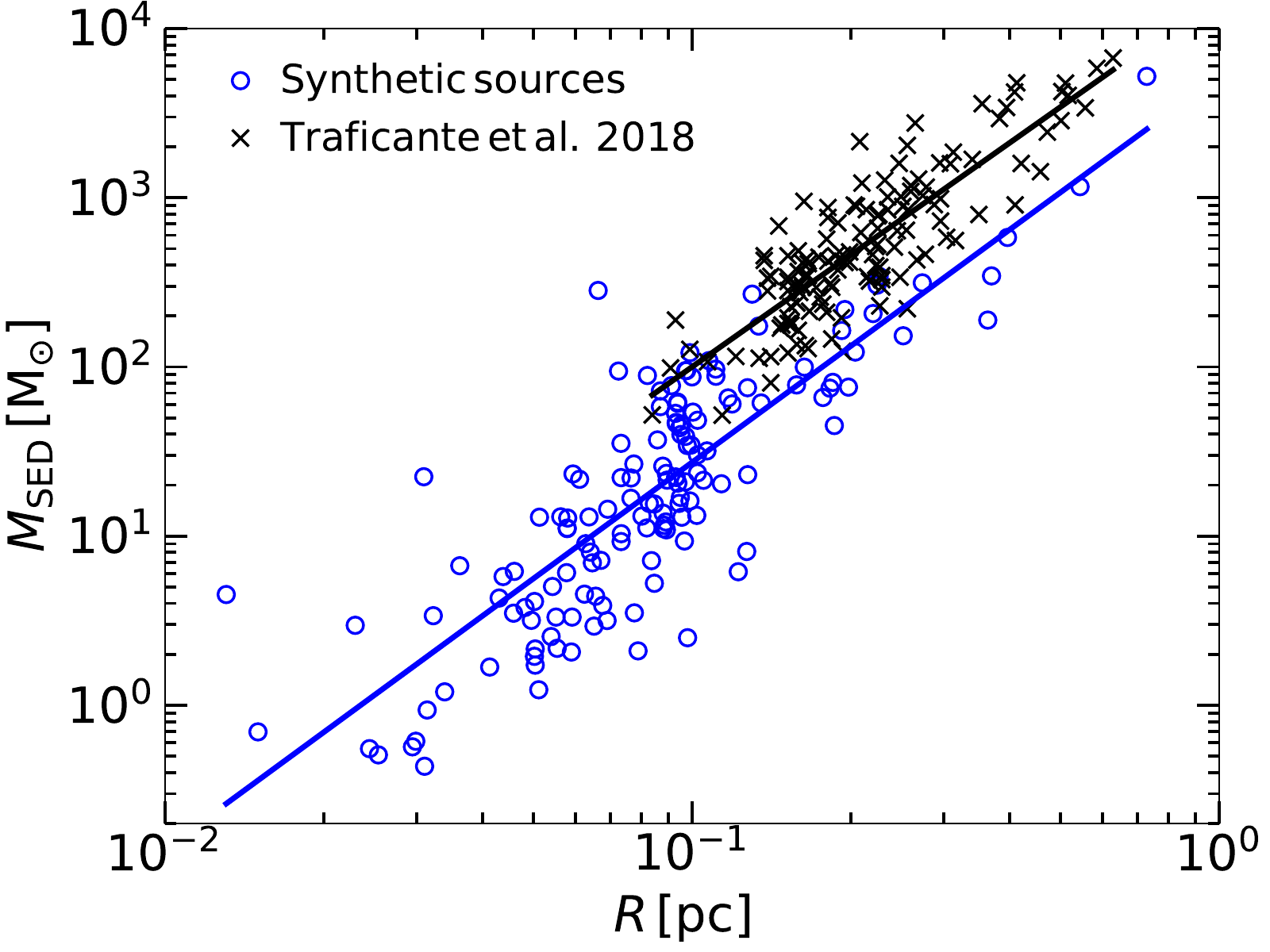}
\caption{Mass versus radius of the compact sources from the synthetic catalog (blue circles) and from the Hi-GAL sources in \citet{Traficante+18}, after the exclusion of HII regions and sources closer than 1.5~kpc (black crosses). The blue and black solid lines are the least-squares fits of the synthetic and Hi-GAL sources, respectively.
Pearson’s $r$ coefficients are 0.84 and 0.85 for the synthetic and observed sources, respectively.}
\label{fig_mass_vs_radius}
\end{figure}

Figure~\ref{fig_mass_vs_radius} shows the relation between the radius, $R$, and the mass, $M_{\rm SED}$, of the synthetic sources in our catalog (blue circles) and of the Hi-GAL source from \citet{Traficante+18} (black crosses). 
The blue and black lines are the linear least-squares fits for the two samples,
\begin{equation}
M_{\rm SED} = 5.235^{+1.966}_{-1.429} \times 10^3 \, \left(\frac{R}{1 \,\rm pc} \right)^{2.28\pm0.13} \, \rm M_{\rm \odot}
\label{eq_m_r_syn}
\end{equation}
for the synthetic sources, and 
\begin{equation}
M_{\rm SED} = 1.579^{+0.327}_{-0.271} \times 10^4 \, \left(\frac{R}{1 \,\rm pc} \right)^{2.20 \pm 0.12} \, \rm M_{\rm \odot}
\label{eq_m_r_obs}
\end{equation}
for the Hi-GAL sources. 

The mass-radius relations from the two samples have the same slopes within the uncertainties, corresponding to almost constant surface density over approximately two orders of magnitude in radius for the synthetic sources and one order of magnitude for the Hi-GAL sources. The surface density of the Hi-Gal sources is on average approximatly three times larger than that of the synthetic sources. This difference in surface density is expected when comparing our synthetic catalog with sources from the Inner Galaxy, as those in \citet{Traficante+18}, while the surface density is essentially the same when comparing with Hi-GAL sources from the Outer Galaxy, as explained in \citet{Lu+2021}.

\subsection{Linewidth-Size Relation}

\begin{figure}
\centering
\includegraphics[width=0.48\textwidth]{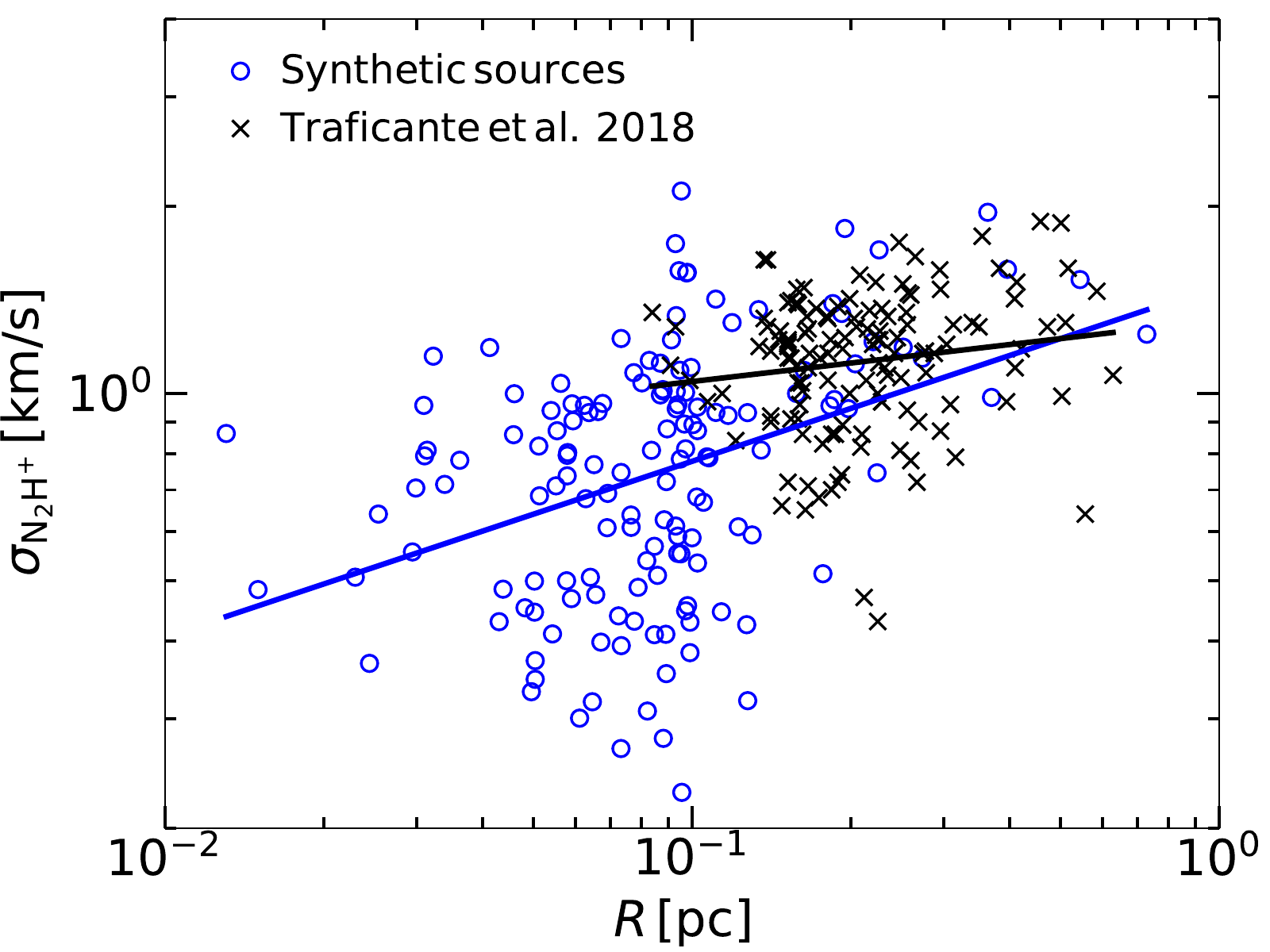}
\caption{
Velocity dispersion of the compact sources estimated from the $\rm N_{2}H^{+}$ spectra as a function of the source radius. Symbols and solid lines are like in Figure~\ref{fig_mass_vs_radius}. 
Pearson’s $r$ coefficients are 0.37 and 0.15 for the synthetic and observed sources, respectively.}
\label{fig_sigma_vs_radius}
\end{figure}

Figure~\ref{fig_sigma_vs_radius} shows the velocity dispersion, $\sigma_{\rm N_{2}H^{+}}$, based on the $\rm N_{2}H^{+}$ lines, as a function of radius, $R$, for the sources in the two samples. The correlation between velocity dispersion and radius is weak for the Hi-GAL sources (Pearson's $r$ coefficient is 0.15), and moderate for the synthetic sources ($r=0.37$).
The power-law fits are:
\begin{equation}
\sigma_{\rm N_{2}H^{+}} = 1.49^{+0.24}_{-0.21} \, \left(\frac{R}{1 \,\rm pc} \right)^{0.28 \pm 0.06} \,\rm km \,\rm s^{-1}
\label{eq_v_r_syn}
\end{equation}
for the synthetic sources, and 
\begin{equation}
\sigma_{\rm N_{2}H^{+}} = 1.31^{+0.12}_{-0.11} \, \left(\frac{R}{1 \,\rm pc} \right)^{0.10 \pm 0.06} \,\rm km \,\rm s^{-1}
\label{eq_v_r_obs}
\end{equation}
for the Hi-GAL sources. 

The slope of the least-squares fit is larger for the synthetic sources, but it is also almost consistent with that of the Hi-GAL sources within the 1-$\sigma$ uncertainties. The steeper slope and stronger correlation may be simply a result of the much larger range of values of source radii in our synthetic sources (approximately two orders of magnitude) than in the observational sample (less than one order of magnitude). In the range of radii where the two samples overlap, it can be seen in Figure~\ref{fig_sigma_vs_radius} that both the median values and the scatter of the velocity dispersion in the two samples are comparable, which is a significant achievement of the simulation, because we have no way to adjust the velocity dispersion through free-parameters to match the observations. The velocity field in the simulation is solely controlled by the SN explosions that are self-consistently modeled by resolving the formation of individual massive stars.

\subsection{The Virial Parameter} \label{obs_alpha_vir}

\begin{figure}
\centering
\includegraphics[width=0.48\textwidth]{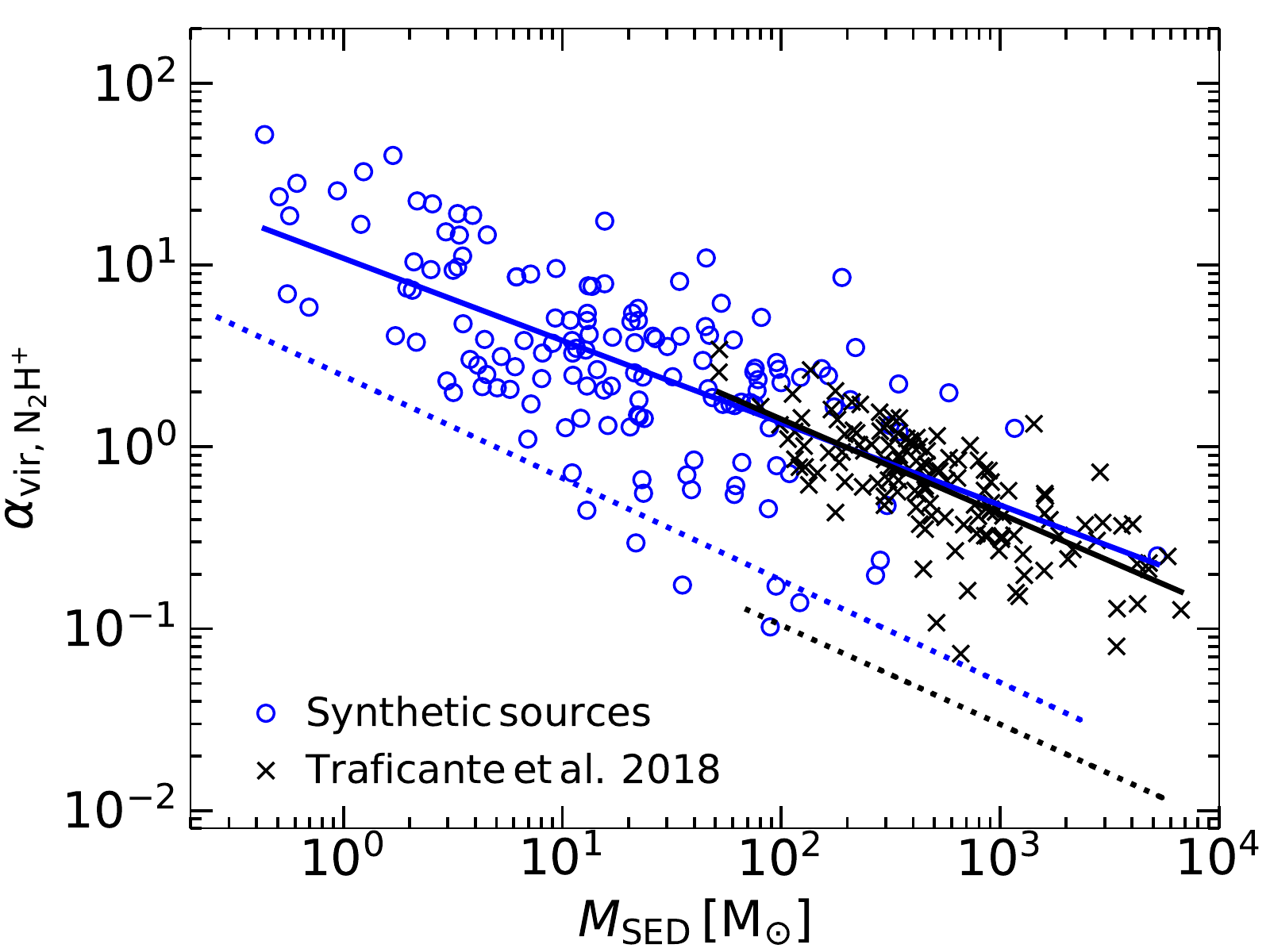}
\caption{Source virial parameter as a function of source mass. Symbols and solid lines are like in Figure~\ref{fig_mass_vs_radius}. The dotted lines show the values of $\alpha_{\rm vir,N_{2}H^{+}}$ as a function of $M_{\rm SED}$ derived from the
least-squares fits of the mass-size relation in Figure~\ref{fig_mass_vs_radius} and adopting a constant velocity dispersion of 0.3~km$s^{-1}$, that is the velocity-channel width used to fit the line profiles. 
Pearson’s $r$ coefficients are -0.64 and -0.73 for the synthetic and observed sources, respectively.}
\label{fig_virial_vs_mass}
\end{figure}

The virial parameter, introduced by \cite{Bertoldi+McKee1992ApJ...395..140B}, measures the ratio of kinetic to gravitational energy, so it is an important indicator of the dynamical state of the compact sources. Its values for all the synthetic and Hi-GAL sources, computed using the expression for a uniform sphere,
\begin{equation}
\alpha_{\rm vir,N_{2}H^{+}} \equiv \frac{5 \sigma^2_{\rm N_{2}H^{+}} R}{G M_{\rm SED} } \, ,
\label{eq_alpha_vir}
\end{equation}
are plotted in Figure~\ref{fig_virial_vs_mass} versus the source mass. 

In the range of masses of the Hi-GAL sources, $\sim$50-7000 $M_{\odot}$, the synthetic and the Hi-GAL sources have  approximately the same mean values and scatter of $\alpha_{\rm vir,N_{2}H^{+}}$ at every mass. The power-law fits for the two samples have the same slopes, within the uncertainty, even when fitting the whole range of masses of our synthetic catalog, $\sim$0.1-6000 $M_{\odot}$, that is nearly five orders of magnitude in mass:
\begin{equation}
\alpha_{\rm vir,N_{2}H^{+}} = 10.90^{+1.79}_{-1.54} \, \left( \frac{M_{\rm SED}}{1 \,\rm M_{\rm \odot}} \right)^{-0.45 \pm 0.05}
\label{eq_vir_r_syn}
\end{equation}
for the synthetic sources, and 
\begin{equation}
\alpha_{\rm vir,N_{2}H^{+}} = 15.56^{+4.76}_{-3.64} \, \left( \frac{M_{\rm SED}}{1 \,\rm M_{\rm \odot}} \right)^{-0.52 \pm 0.04}
\label{eq_vir_r_obs}
\end{equation}
for the Hi-GAL sources.

Given the poor correlation between source velocity dispersion and size, this dependence of $\alpha_{\rm vir,N_{2}H^{+}}$ on $M_{\rm SED}$ is primarily a consequence of the mass-size relation. Assuming a single characteristic value for the velocity dispersion of all sources yields a relation between virial parameter and mass consistent with that found for the compact sources. In Figure~\ref{fig_mass_vs_radius}, the dotted lines show the prediction for a constant velocity dispersion equal to the velocity channel width of the $\rm N_{2}H^{+}$ spectra after velocity smoothing, 0.3~km s$^{-1}$, which is considered as a lower limit for the velocity dispersion that can be measured in the sources with those spectra. The lines are expected to mark the lower envelope of the scatter plots, as they approximately do. This shows that the absence of compact sources of both low $\alpha_{\rm vir,N_{2}H^{+}}$ and low $M_{\rm SED}$ is at least partly a consequence of the velocity resolution of the spectral lines.

\section{Comparison with the 3D MHD Simulation} \label{comparison_mhd}

Relations derived in the previous section apply to compact sources with mass, size and velocity dispersion estimated from dust continuum and molecular line observations. In the following, we use the simulation datacubes to identify the 3D clumps corresponding to the compact sources, and derive their statistical relations. For each source, we consider the 250 pc gas column whose projection along the line of sight of the synthetic observations corresponds to that source, with the gas density averaged in each 2D slice of the source area. We then search for the highest density along the gas column, and define the main clump corresponding to the compact source as the sphere of radius equal to the source radius, $R$, centered around that density peak. The clump mass, $M_{\rm main}$, and velocity dispersion, $\sigma_{\rm main}$, are both measured summing over all the computational cells inside that sphere. We refer to the clump as the {\it main} clump, as other secondary density peaks may exist along the same line of sight. 

A clump identified with this procedure may be part of a larger density enhancement, such as a larger clump or a filament, with a significant extension on the plane of the sky or along the line of sight. Thus, our clump sample may differ significantly from a sample selected directly with a 3D search algorithm, such as a dendrogram analysis \citep{Rosolowsky+08}, where neither the shape nor the size of a clump is constrained by properties related to its 2D projection. Our analysis is also different from previous works where 3D search algorithms are used to compare 3D density structures with position-position-velocity structures \citep[e.g.][]{Beaumont+13}, as our source identification has been carried out on the 2D dust-continuum maps, like in the Hi-GAL observations, while the velocity information is only used to exclude sources after the initial selection. However, our clump definition is appropriate for the purpose of testing observational results based on the interpretation of compact sources as spherical clumps of size equal to the source size, which is the goal of this work. This spherical clump assumption is used, for example, in the calculation of the observed virial parameter, $\alpha_{\rm vir}$ (see \S~\ref{obs_alpha_vir}.)

To check for the mass uncertainty related to the spherical assumption, we have also tested an alternative definition of the 3D clump masses. We cannot consider the possibility of a 3D clump volume whose projection extends outside of the area covered by the corresponding 2D source, because that extra volume would not contribute to the luminosity of the source. However, we can at least consider the case of clumps elongated in the direction of the line of sight, with a length larger than the source diameter. Thus, in the alternative definition of the 3D clump mass, we define the clump as confined by the source radius on the plane of the sky, and by the local minima around the maximum density peak along the line of sight. This procedure results in 3D masses that are on average only 3\% larger than in the spherical approximation. However, the relation between these masses and the spherical ones is shallower than linear, meaning that the masses tend to be overestimated for low-mass 3D clumps, and underestimated for high-mass 3D clumps. As a result, the standard deviation of the relative mass differences is approximately 40\%. In the following (see \S~\ref{mass-versus-mass}), we will show that the mass of the synthetic sources tends to overestimate that of their corresponding main clumps at large masses, and underestimate them at low masses (see Figure~\ref{fig_mass_comparison}). Thus, if the alternative definition of the main clump mass were adopted, the discrepancy between the masses of the synthetic sources and those of the main clumps would be even larger.

\subsection{Mass and Velocity Dispersion of Sources versus 3D Clumps} \label{mass-versus-mass}

\begin{figure}
\centering
\includegraphics[width=0.48\textwidth]{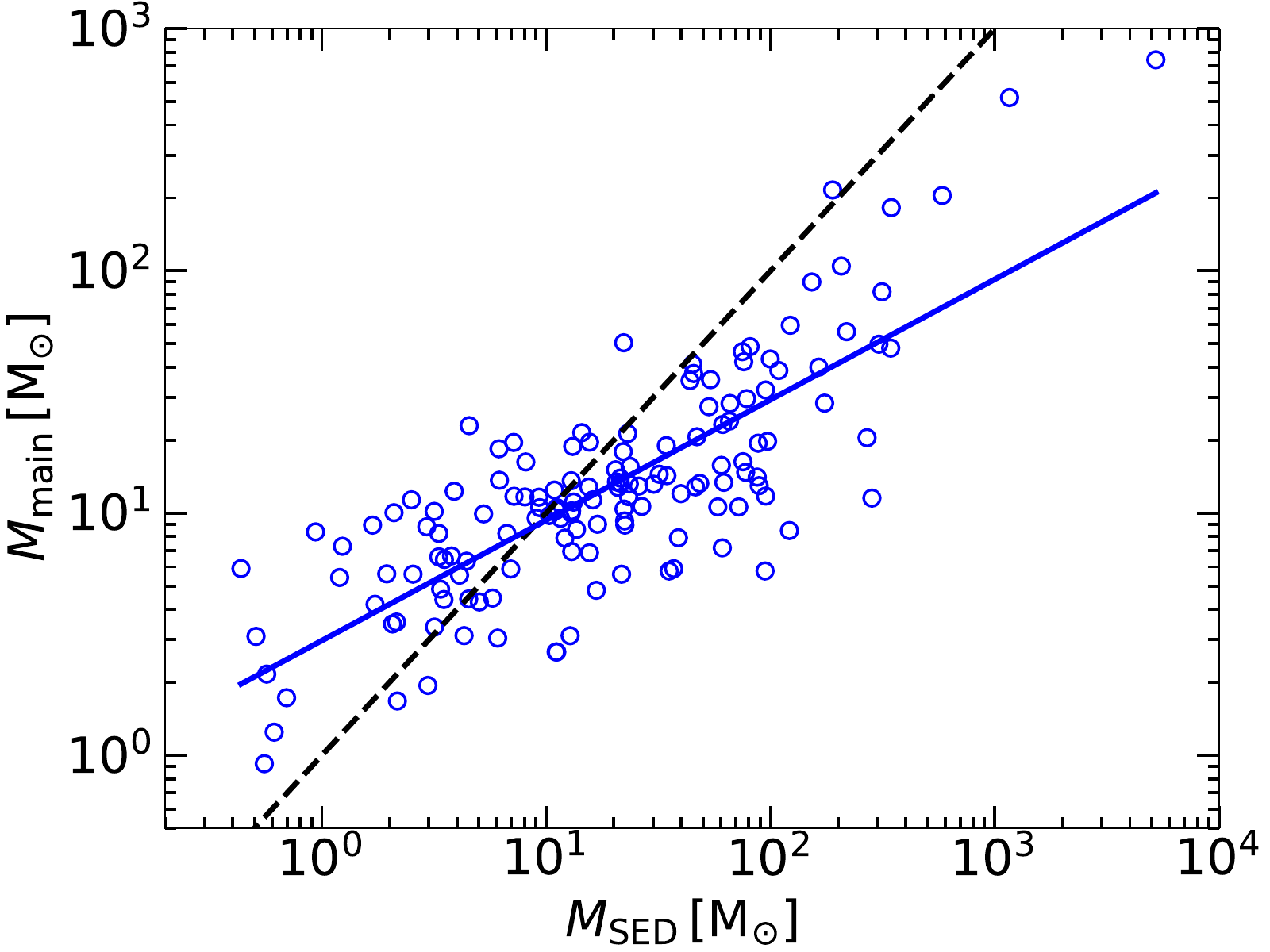}
\caption{The mass of main 3D clumps from the MHD simulation, $M_{\rm main}$, as a function of the mass of the corresponding compact source derived from the SED fitting, $M_{\rm SED}$. The black dashed line is one-to-one line and the solid blue line the least-squares fit giving $M_{\rm main}=2.976^{+0.338}_{-0.304} \, M_{\rm SED}^{0.50 \pm 0.03}$.
Pearson's $r$ coefficient is 0.79.}
\label{fig_mass_comparison}
\end{figure}

In Figure~\ref{fig_mass_comparison}, we compare the mass, $M_{\rm main}$, of the main 3D clump along the line of sight with the mass, $M_{\rm SED}$, of the corresponding compact source. The least-squares fit shown by the solid blue line has a rather shallow slope of 0.50, $M_{\rm main} \propto M_{\rm SED}^{0.50}$, meaning that the estimated mass of the compact sources tends to increasingly overestimate the mass of the corresponding clumps towards larger masses, and to underestimate it towards lower masses. The two masses are approximately comparable on average at around 10~$\rm M_{\odot}$, while $M_{\rm SED}$ can be 10 times larger than $M_{\rm main}$ at around 100~$\rm M_{\odot}$, or 10 times smaller for values around 1~$\rm M_{\odot}$. At high masses, $M_{\rm SED}$ is on average 2-3 times higher than $M_{\rm main}$, but slightly smaller than the total mass in the line of sight. Thus, some of this mass mismatch can be explained simply by projection effects: $M_{\rm SED}$ traces the total mass in the line of sight rather than that of an individual 3D clump. However, the same cannot be said for cases where $M_{\rm SED}$ is even larger than the total mass, or at smaller masses, where $M_{\rm SED}$ is smaller than $M_{\rm main}$. 
These additional discrepancies between $M_{\rm main}$ and $M_{\rm SED}$, already noticed in \citet{Lu+2021}, are specific to Hi-GAL's band-merging procedure, and are more complex to grasp than in the case of aperture photometry. As our source catalog is created by following Hi-GAL's source extraction and photometry in great detail in \citet{Lu+2021}, we are confident that these mass discrepancies apply to High-GAL sources as well. Because the differences between $M_{\rm main}$ and $M_{\rm SED}$ are a key result of this work, we dedicate Appendix~\ref{app_band_merging} to clarify how they arise from the band-merging procedure.

\begin{figure}
\centering
\includegraphics[width=0.48\textwidth]{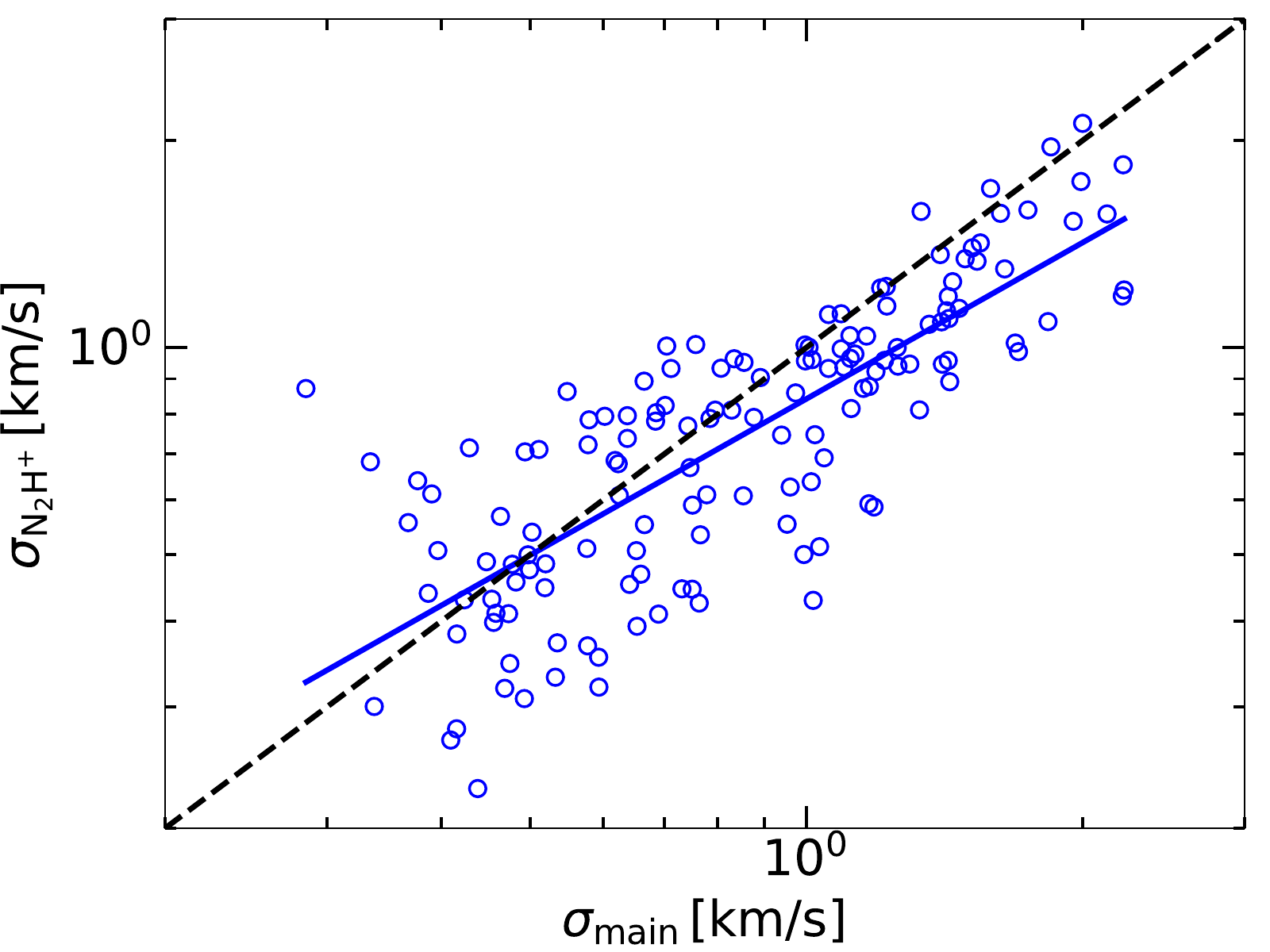}
\caption{The comparison of the velocity dispersion of the main clumps from the MHD simulation and the velocity dispersion of the corresponding synthetic sources from the $\rm N_{2}H^{+}$ line profiles. The blue solid line is the least-squares fit to the data points giving $\sigma_{\rm N_{2}H^{+}}=0.84^{+0.02}_{-0.02} \, \sigma_{\rm main}^{0.75 \pm 0.05}$, and the black dashed line is the one-to-one line.
Pearson’s $r$ coefficient is 0.79.}
\label{fig_sigma_comparison}
\end{figure}

For the calculation of the velocity dispersion of the main 3D massive clumps along the light of sight, we adopt the one-dimensional rms velocity of the computational cells within the clumps,
\begin{equation}
\sigma_{\rm main} \equiv  \left[{\frac{1}{3N}}\,\sum_ {i=1}^{3} \, \sum_{n=1}^{N} ( u_{i,n} - \bar{u}_i )^2  \right ]^{1/2} \, ,
\label{eq_v}
\end{equation}
where N is the total number of cells within the main clump, $\bar{u}_i \equiv \sum_{n=1}^{N} u_{i,n} /N$ are the components of the mean velocity of the cells, and $u_{i,n}$ is the $i$-th velocity component of the $n$-th cell. 

In Figure~\ref{fig_sigma_comparison}, we compare the velocity dispersion derived from the main clumps, $\sigma_{\rm main}$, with the velocity dispersion estimated from the synthetic $\rm N_{2}H^{+}$ line of the corresponding sources, $\sigma_{\rm N_{2}H^{+}}$. The two estimates of the velocity dispersion are strongly correlated, and directly proportional to each other on average, with a least-squares fit with slope 0.8 (blue solid line). Despite the good correlation, the total scatter in the plot is still significant, around a factor of 2 or 3, which is to be expected if the random velocity field in the clumps is not isotropic, as discussed in \S~\ref{discussion}. The strong correlation in Figure~\ref{fig_sigma_comparison} demonstrates that, despite the projection and confusion effects, the $\rm N_{2}H^{+}$ emission lines tend to trace primarily the kinematics of the main clump in the line of sight. This is hardly surprising because of the relatively high critical gas density for the excitation of the lines, and given that the line is optically thin, so it probes the whole clump mass.

\subsection{Mass-Size Relation}

\begin{figure}
\centering
\includegraphics[width=0.48\textwidth]{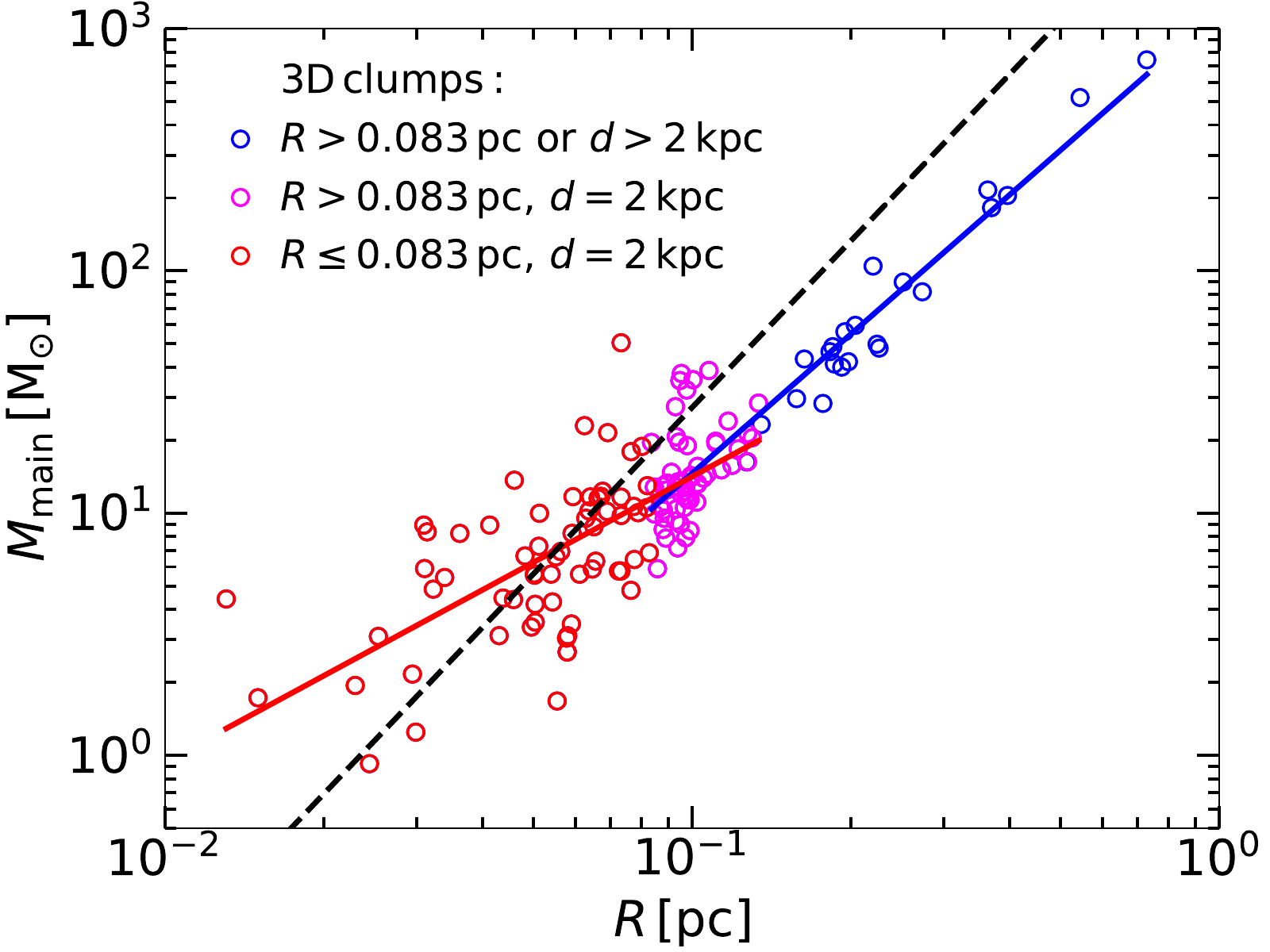}
\caption{Mass-size relation of the main 3D clumps. The solid blue line is the least-squares fit for clumps with $R > 0.083$~pc (blue and magenta circles) or distance ($d$) larger than 2~kpc (blue circles), and the solid red line is the fit for clumps at a distance of 2~kpc (magenta and red circles). The dashed black line is the fit to the $M_{\rm SED}$-$R$ relation of the synthetic sources from Fig.~\ref{fig_mass_vs_radius}. 
Pearson’s $r$ coefficient is 0.87.}
\label{fig_mass_vs_radius_mhd}
\end{figure}

Figure~\ref{fig_mass_vs_radius_mhd} shows the mass-size relation for the main clumps. The black dashed line is the least-squares fit to the mass-size relation of the compact sources, shown in Figure~\ref{fig_mass_vs_radius}. 
For radii larger than 0.1~pc, almost all points lie below the dashed line, while for radii smaller than 0.05~pc, almost points are above the line. Because the slope of the mass-size relation seems to change around 0.1~pc, we have computed the least-squares fit for two samples separately: the clumps associated to sources found at a distance of 2~kpc (red and magenta circles), 
\begin{equation}
M_{\rm main} = 2.125^{+0.697}_{-0.525} \times 10^{2} \, \left(\frac{R}{1 \,\rm pc} \right)^{1.18\pm0.11} \, \rm M_{\rm \odot} \, ,
\label{eq_m_r_clumps}
\end{equation}
and those with diameter larger than Herschel's 250~$\mu$m beam size at 2~kpc, which is 0.083~pc (magenta and blue circles), or at a distance larger than 2~kpc (blue circles),
\begin{equation}
M_{\rm main} = 1.181^{+0.239}_{-0.199} \times 10^{3} \, \left(\frac{R}{1 \,\rm pc} \right)^{1.91\pm0.09} \, \rm M_{\rm \odot}\,.
\label{eq_m_r_clumps_large}
\end{equation}
While the span of mass values for the larger clumps is due primarily to the span in the distances of the associated sources (from 2 to 12~kpc), the span in masses of the smaller clumps is due to the range of values of the radii of the sources at 2~kpc, which is the result of the beam-deconvolution of the source radii when these are resolved, meaning they are larger than the beam (see \S~\ref{sample} for the computation of the source radius).     

At the smaller radii, the power-law fit has a slope of 1.2 (the red line in Figure~\ref{fig_mass_vs_radius_mhd}), while the slope at the larger radii is 2.0 (the blue line), that is consistent with the surface density being independent of clump size on average. The nearly constant surface density is an intrinsic properties of the larger clumps, rather than a trivial consequence of our density selection. Our minimum density threshold (see \S~\ref{sample}) scales with the square of the distance, hence the square of the clump size, which would yield a surface density decreasing with clump size, rather than constant. The slope of 2.0 is nearly the same as that of the mass-size relation of the compact sources, but the least-squares fit lies a factor of 2-3 below that of the sources, meaning that, at the larger radii, the clumps are on average a factor of 2-3 less massive than the corresponding sources. 

The shallower slope of the mass-size relation of the clumps at smaller radii, relative to the mass-size relation of their associated compact sources, can be understood as the effect of the shallow slope of the $M_{\rm main}-M_{\rm SED}$ relation (see Figure~\ref{fig_mass_comparison}). It shows that, at a fixed distance (2~kpc) more compact clumps tend to have a larger surface density than larger ones, a trend that is masked in the case of the compact sources, because the smaller ones tend to underestimate the clump mass, as explained in \S~\ref{mass-versus-mass}.

\subsection{Linewidth-Size Relation}

\begin{figure}
\centering
\includegraphics[width=0.48\textwidth]{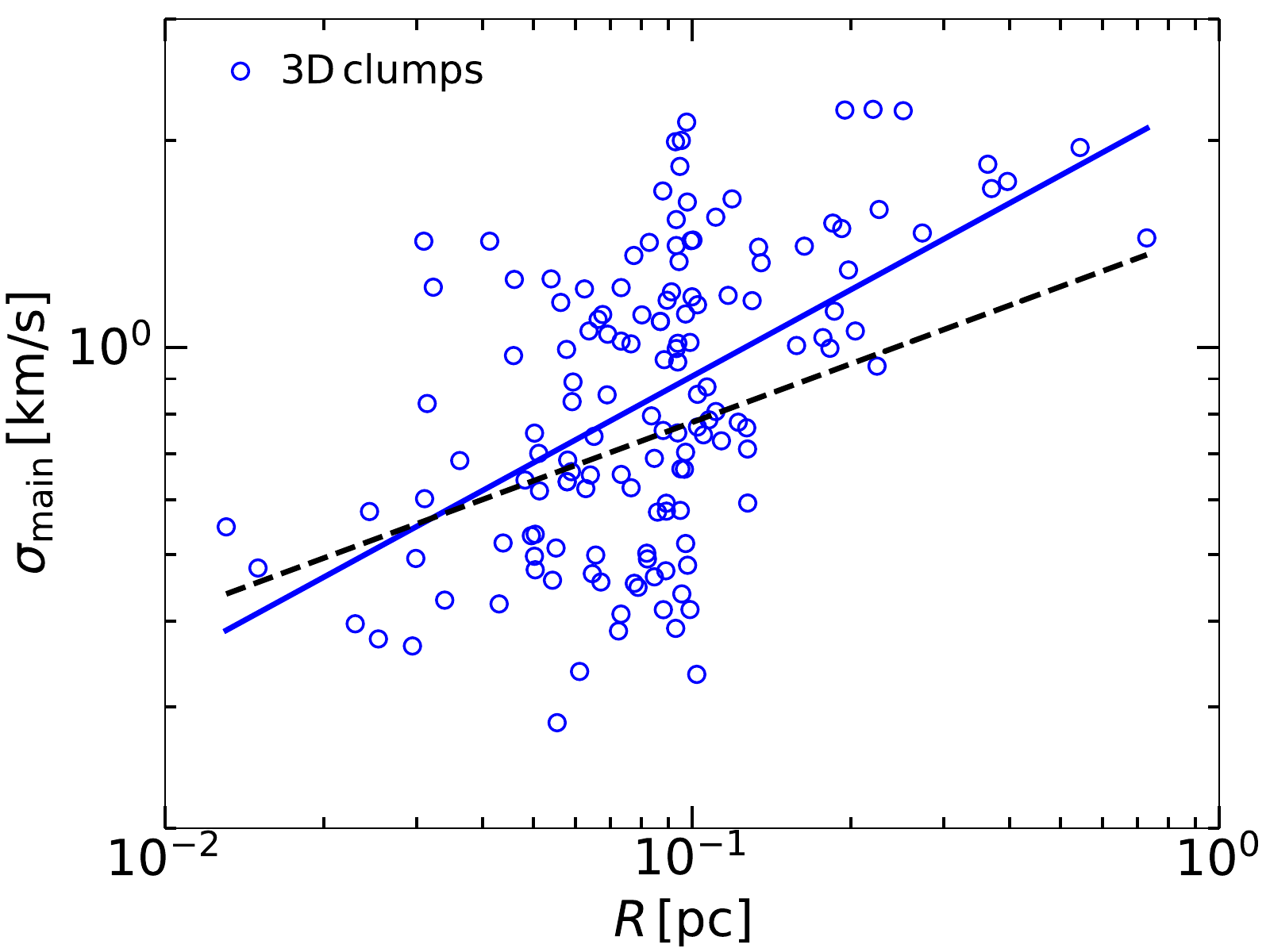}
\caption{Velocity-size relation of the main 3D clumps. The solid blue line is the least-squares fit to the data points, and the black dashed line is the fit to the velocity-size relation of the synthetic sources from Fig.~\ref{fig_sigma_vs_radius}. Pearson’s $r$ coefficient is 0.53.}
\label{fig_sigma_vs_radius_mhd}
\end{figure}

In Figure~\ref{fig_sigma_vs_radius_mhd}, we plot the relation between the velocity dispersion in the main clumps, $\sigma_{\rm main}$, and their radius, $R$. The solid blue line is the least-squares fit to the points, giving
\begin{equation}
\sigma_{\rm main} = 2.37^{+0.37}_{-0.32} \, \left( \frac{R}{1\,\rm pc} \right)^{0.42\pm0.06} \,\rm km \,\rm s^{-1}
\label{eq_v_r_mhd}
\end{equation}
Within the 1-$\sigma$ uncertainties, both the median values of the velocity dispersion and the slope of this fit are nearly consistent with those of the compact sources (dashed line in Figure~\ref{fig_sigma_vs_radius_mhd}), as expected from the good correlation between the velocity dispersion estimated from the N$_{2}$H$^+$ line and that of the clumps, already shown in Figure~\ref{fig_sigma_comparison}. This shows that the N$_2$H$^+$ lines are able to probe well the velocity field within the clumps, at least within the boundaries of the deconvolved radius of the Hi-GAL sources.  

Despite the similarities between the velocity dispersion of the sources and the clumps, the correlation between velocity dispersion and size in the case of the clumps is a bit stronger than for the synthetic sources (Perason's $r=0.53$ and 0.37, respectively). The weakening of the correlation, as well as the reduction in the slope of the power-law fit, in going from the 3D clumps to the sources can be understood as the result of random differences in the velocity dispersion between the 1D and 3D measurements, as mentioned below in \S~\ref{discussion}.

\subsection{The Virial Parameter}

\begin{figure}
\centering
\includegraphics[width=0.48\textwidth]{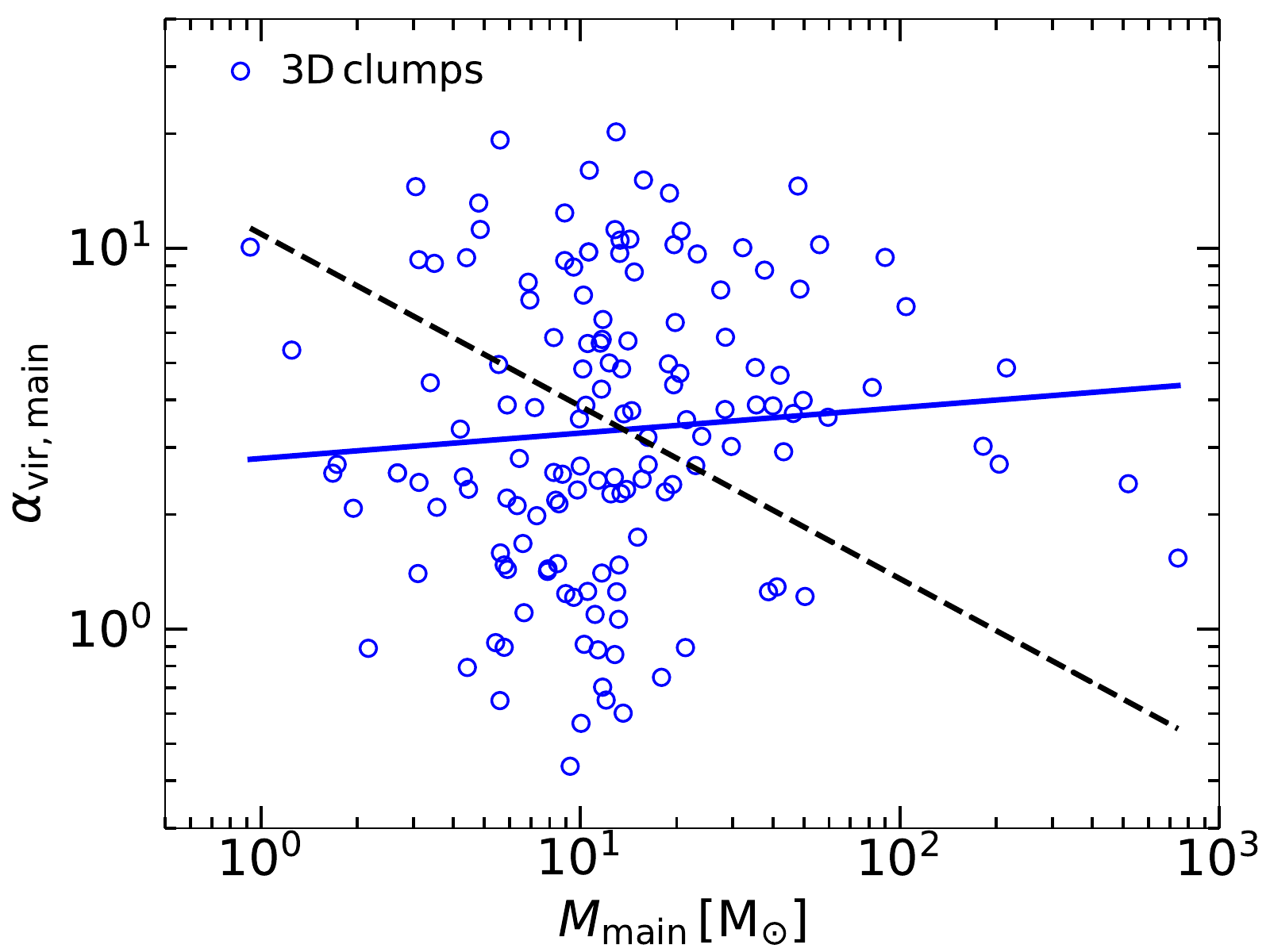}
\caption{The relation between virial parameter and mass of the main 3D clumps. The solid blue line is the least-squares fit to the data points, and the dashed line is the fit to the same relation for the synthetic sources from Fig.~\ref{fig_virial_vs_mass}. Pearson’s $r$ coefficient is 0.08.}
\label{fig_virial_vs_mass_mhd}
\end{figure}

The virial parameter of the 3D clumps is computed as the ratio of their kinetic and gravitational energies, where the kinetic energy is computed as the sum of the kinetic energy of the computational cells inside the spheres that define the clumps, while the gravitational energy is approximated by the formula for a uniform sphere:
\begin{equation}
\alpha_{\rm vir,main} \equiv 2 \frac{E_{\rm K}}{E_{\rm G}} \equiv 
\frac{\sum_ {i=1}^3 \, \sum_{n=1}^N m_{n}( u_{i,n} - \bar{u}_{i} )^2 }{3 G M^2_{\rm MHD}/5R } \, .
\label{eq_alpha_vir_mhd}
\end{equation}

The relation between the virial parameter of the 3D clumps, $\alpha_{\rm vir,main}$, and their mass, $M_{\rm main}$, is shown in Figure~\ref{fig_virial_vs_mass_mhd}, where the solid blue line is the least-squares fit to the points, given by
\begin{equation}
\alpha_{\rm vir,main} = 2.81^{+0.59}_{-0.49} \,\left( \frac{M_{\rm main}}{1\,\rm M_{\odot}} \right) ^{0.07\pm0.07} \, ,
\label{eq_vir_m_mhd}
\end{equation}
and the dashed black line the fit to the corresponding relation for the compact sources (see Figure~\ref{fig_virial_vs_mass}). There is essentially no correlation between the virial parameter and the mass of the main clumps (Pearson's $r=0.08$). This is in sharp contrast with the clear decrease in virial parameter with increasing mass for the associated compact sources shown by Figure~\ref{fig_virial_vs_mass} (where $r=-0.64$). It seems that, in going from the 3D clumps to the synthetic observations, the correlation is artificially generated by the fact that the range in the source masses is stretched relative to that of the clumps, because the source masses overestimate the clump masses at large values, and underestimate them at low values, as discussed in \S~\ref{mass-versus-mass}.

As explained in the context of Figure~\ref{fig_virial_vs_mass}, the relation between virial parameter and mass of the sources is primarily the result of their mass-size relation (essentially their surface density does not depend on size on average), given the weak size dependence of their velocity dispersion. In the case of the clump, the surface density is independent of size only for the larger clumps, as shown in Figure~\ref{fig_mass_vs_radius_mhd}, and in fact, above approximately 60~$\rm M_{\odot}$, the clump virial parameter shows a trend of decreasing with increasing mass, particularly in the upper envelope of the scatter plot. However, due to the lower masses and slightly larger velocity dispersion relative to the sources, the most massive clumps have a significantly larger virial parameter than the most massive sources. The average value for the 10 most massive clumps is $\alpha_{\rm vir,main}=4.9$, while for the 10 most massive compact sources it is $\alpha_{\rm vir,N_{2}H^{+}}=1.3$. Thus, while the most massive compact sources appear to have rather low virial parameters, consistent with being gravitationally bound on average, their corresponding 3D clumps have generally an excess of kinetic energy that would classify them as unbound. The converse is also true: the median mass of the sources with $\alpha_{\rm vir,N_{2}H^{+}}<2$ is rather large, 61~$\rm M_{\odot}$ (to be compared with the median mass of the sources with $\alpha_{\rm vir,N_{2}H^{+}}>2$, which is 11~$\rm M_{\odot}$), while the median mass of the clumps with $\alpha_{\rm vir,main}<2$
is only 10~$\rm M_{\odot}$ (even smaller than the median mass of the clumps with $\alpha_{\rm vir,main}>2$, which is 13~$\rm M_{\odot}$).

\section{Discussion} \label{discussion}

The main purpose of deriving statistical relations for Hi-GAL compact sources is to establish the dynamical state of massive clumps, improving our understanding of their origin, evolution and star-formation process (we focus on protostellar clumps). In \citet{Traficante+18}, the low values of the virial parameter of the most massive sources is taken as evidence that their internal motions are a signature of gravitational collapse, rather than turbulence. The large range of virial parameter of their sources, though, shows that gravitational collapse could at best explain a minor fraction of the sources' velocity dispersion. If anything, it could explain the sources with masses around 100~$\rm M_{\odot}$, whose virial parameter is around 2, the value expected for free-fall motions. The most massive sources in \citet{Traficante+18}, between $10^3$ and $10^4$~$\rm M_{\odot}$, have a virial parameter on average approximately 10 times smaller, which begs the fundamental question of how can massive clumps exist with such low virial parameters. Such clumps should be gravitationally unstable, hence free-fall motions (virial parameter $\sim 2$) should develop within a free-fall time, that is $\lesssim 10^5$~yr, at the characteristic mean density of $\sim 10^5$~cm$^{-3}$ of these massive clumps.

Our results significantly alleviate this conundrum, because we have found that the systematic drop in virial parameter of the synthetic sources with increasing mass is largely an observational artifact, whose origin lies primarily in the incorrect estimate of the clump mass \citep[see][for different observational artifacts\footnote{The observational effects discussed in \citet{Singh+2021} do not apply to the way the source parameters are derived here. \citet{Traficante+18bias} propose that the velocity dispersion from the N$_2$H$^+$ line does not sample the whole clump, but we have demonstrated here that it actually retrieves the clump velocity dispersion well, so that argument does not apply here either.}]{Traficante+18bias,Singh+2021}. Based on our results, the real virial parameters of the actual 3D clumps associated to the most massive Hi-GAL sources may easily be several times larger than what is inferred from the observations. In addition, we have shown in \citet{Lu+2021} that the projection effects that lead to overestimate the mass of the main clumps associated to the massive sources are expected to be larger for sources in the Inner Galaxy than for sources in the Outer Galaxy and in our synthetic catalog. Thus, the needed correction to the virial parameters of the most massive sources in \citet{Traficante+18} may be even larger than in the case of our synthetic sources.  

Significant discrepancies between the virial parameters of 3D clumps and their observational counterparts have been documented in previous works based on synthetic observations of simulations \citep[e.g.][and references therein]{Beaumont+13}, though not as large as in this work. The comparison is often carried out by using the same algorithm to extract connected structures in position-position-position (PPP) space (from the density cube of a simulation) and in position-position-velocity (PPV) space, from synthetic observations of molecular emission lines. Our source masses are derived from the dust-continuum maps and therefore represent the total mass in the line of sight (minus the background). In \citet{Beaumont+13}, the source masses are from the PPV analysis of CO lines, so the mass in the line of sight can be broken into different components. This may result in a better match of the clumps in PPP space than using 2D dust maps.
Furthermore, in our work, the size of the simulated volume allows a projection over 250~pc, over 10 times larger than in \citet{Beaumont+13}. The difference in the analysis method and the larger projection distance may explain why we find even larger mass and virial parameter discrepancies between 2D sources and 3D clumps than in \citet{Beaumont+13}.

While offering a possible solution to the problem of the existence of too many compact sources with too small virial parameters, our results do not support a scenario where gravitational collapse explains the observed velocity dispersion of the compact sources. Even when measured for the actual 3D clumps, the range of virial parameters remains very large, between 0.4 and 20.2, with a median value of 3.2. Neither the median value nor the large scatter are consistent with gravity being the leading cause of the observed motions. The natural explanation, consistent with our simulation, is that the origin of the motions is the large-scale turbulence, whose random converging flows are also the explanation for the origin of the clumps themselves. 

The lack of a strong and positive velocity-size correlation in the compact sources is presented by \citet{Traficante+18}, as well as by other authors finding similar results
\citep[e.g.][]{Caselli+Myers95,Plume+97,Shirley+03,Gibson+09,Kirk+17}, as evidence against the turbulent origin of the observed motions in massive clumps. However, irrespective of the observational details, massive clumps provide an incomplete view of the overall turbulence, strongly biased towards regions of converging flows, where velocity differences are well above the average for that scale. One should not expect to retrieve the velocity scaling representative of the average turbulent cascade from such a strongly biased sampling of the turbulent velocity field. But this does not imply that the turbulent cascade is not the reason for the observed velocity dispersion in the clumps. 

\begin{figure}
\centering
\includegraphics[width=0.48\textwidth]{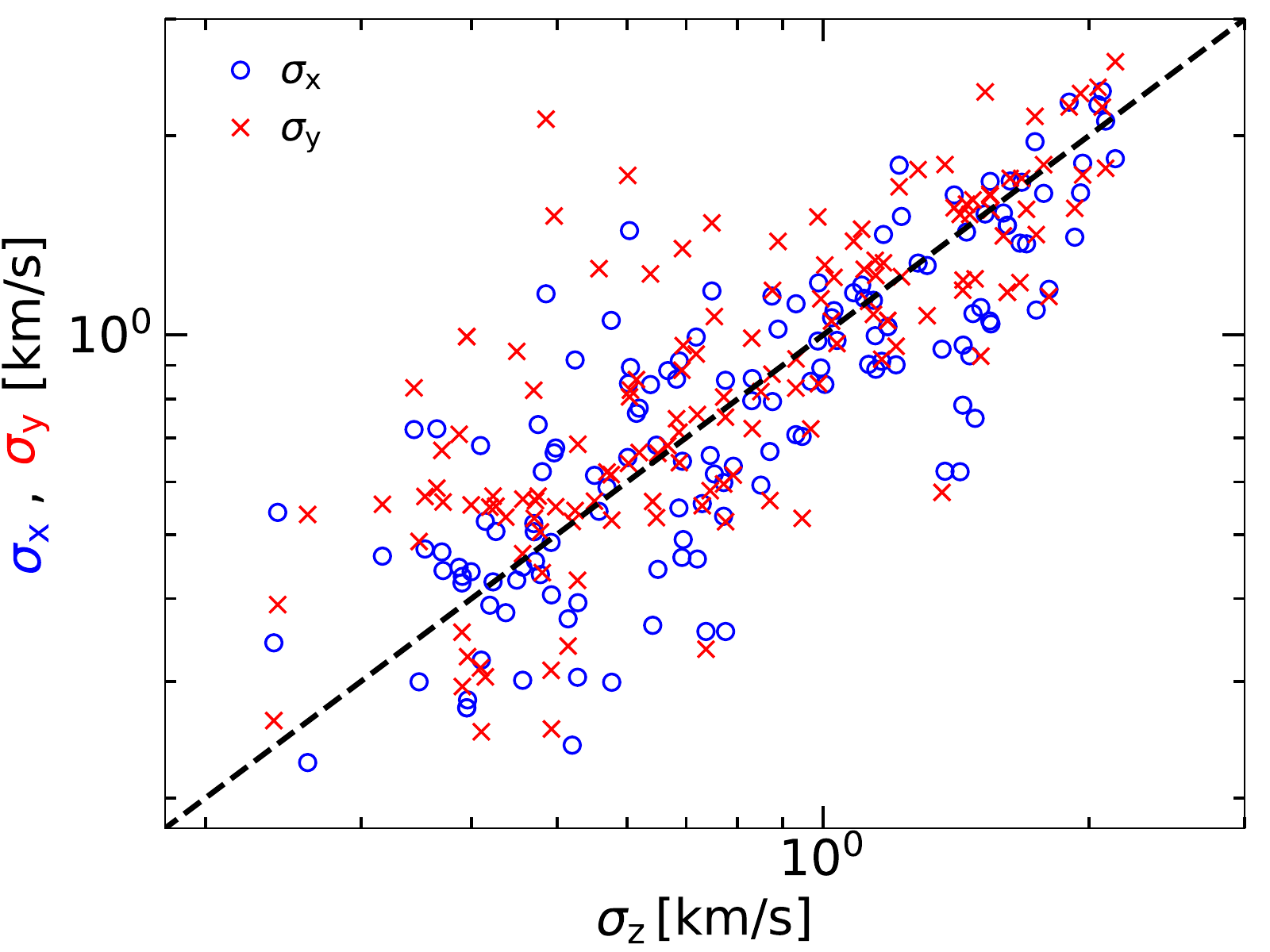}
\caption{Comparison of 1D velocity dispersion of the main 3D clumps measured from different directions. The blue circles and the red cross are the $\sigma_{\rm x}$ - $\sigma_{\rm z}$ and  $\sigma_{\rm y}$ - $\sigma_{\rm z}$ relations, respectively. The black dashed line is the one-to-one line. }
\label{fig_sigma_xyz}
\end{figure}

We also find that the velocity-size correlation is stronger in the case of the 3D clumps than in the case of the synthetic sources, suggesting that the failure of retrieving Larson's velocity-size relation \citep{Larson81} for the compact sources is partly an observational limitation. Figure~\ref{fig_sigma_xyz} shows that there is a significant scatter in the relation between the 1D velocity dispersion of a clump measured in different directions. This is further evidence that the velocity dispersion in massive clumps is the result of random flows converging from larger scales, rather than of more relaxed isotropic turbulent motions within the clumps (hence the very large values of the observed velocity dispersion relative to a characteristic Larson's velocity-size relation at that scale). Because of the large scatter between different directions shown in Figure~\ref{fig_sigma_xyz}, there can be significant differences between the 3D velocity dispersion in a clump, and the 1D velocity dispersion measured in the corresponding compact source. This random error in the velocity dispersion weakens the velocity-size correlation in the compact sources, relative to that of the 3D clumps.

\section{Conclusions} \label{conclusions}

In this work, we have selected a subsample of 145 objects from our much larger catalog of synthetic compact sources presented in \citet{Lu+2021}. The main catalog was computed following exactly the same procedure as the Hi-GAL catalog \citep{Elia+2017,Elia+21}, while the smaller sample studied here was selected based on the N$_2$H$^+$ emission, as in \citet{Traficante+18} and the requirement that the sources are protostellar (as traced by 70~$\mu$m emission). By comparing the properties of the synthetic sources first with those from the real observations, and then with those of their corresponding 3D clumps in the simulation, we have arrived at the following results:       

\begin{enumerate}

\item The mass-size, velocity-size, and virial parameter-mass relations of the synthetic sources have almost the same slope, within the uncertainties, and approximately the same median values and scatter as those of the observed sources, in the ranges of size or mass where they overlap. The only discrepancy is that the surface density of the synthetic sources is approximately three times smaller than the observed sources, as expected when comparing our catalog with sources from the Inner Galaxy.

\item The masses of the compact sources derived from their SEDs overestimate those of the associated 3D clumps at high masses, and underestimate them at low masses. This mismatch results mainly from two band-merging effects that alter the construction of the  SEDs (see Appendix~\ref{app_band_merging}) and from projection effects. Both band-merging and projection effects are expected to influence also the sources in the Hi-GAL catalog, with the projection effects possibly even larger for sources from the Inner Galaxy, like those in the observational sample considered in this work.

\item The velocity dispersion estimated from the N$_2$H$^+$ emission lines of the synthetic sources is on average comparable with the velocity dispersion measured from the 3D velocity field within the corresponding main clumps. However, the 1D velocity dispersion of a clump measured from different directions has significant variations, which explains the weaker correlation between velocity dispersion and size of the sources relative to that of the 3D clumps.   

\item The virial parameter and the mass of the 3D clumps are not correlated, in sharp contrast to the case of the synthetic and observed sources. The correlation in the (synthetic) observations arises primarily from the errors in the source masses, because large masses are overestimated and low masses underestimated. As a result, the most massive clumps do not have the very low values of virial parameters that are measured in the most massive sources.

\item Given the wide range of values ($\sim 0.4-20.2$) and the relatively high median value ($\sim 3.2$) of their virial parameters, the dynamical state of the 3D clumps cannot be primarily the result of self-gravity, even when their associated compact sources have estimated virial parameters consistent with gravitationally bound or free-falling objects. The velocity dispersion of the sources is naturally explained by the presence of random converging flows in the large-scale turbulence that are also responsible for the formation of the clumps. 

\end{enumerate}

The main conclusion of this work is that massive clumps associated with massive Hi-GAL sources may have intrinsic values of both mass and virial parameter that differ significantly from those estimated for the corresponding sources. In particular, massive sources with very low virial parameter would generally correspond to real 3D clumps with significantly lower mass and significantly larger virial parameter. The actual clumps are not necessarily unbound, as the large velocity dispersion originates from random converging flows in the process of dissipating much of their kinetic energy in a complex system of shocks. This results in the turbulent fragmentation of the clumps leading to the formation of stellar clusters, as well documented by an ever increasing number of interferometric studies. In addition to these high-resolution follow-ups of individual massive clumps, future work should further the analysis of very large clump samples like Hi-GAL and MALT90. Their statistical comparison with high dynamic range simulations is necessary to shed light on the process of star formation at large scales and thus on the origin of massive stars.

\section*{Acknowledgements}

The authors are grateful to the anonymous referee for a useful referee report.
We thank Alessio Traficante for his assistance and for providing detailed information about his dataset.
ZJL acknowledges financial support from China Scholarship Council (CSC) under grant No. 201606660003.
PP and VMP acknowledge support by the grant PID2020-115892GB-I00 funded by MCIN/AEI/10.13039/501100011033. 
Computing resources for this work were provided by the NASA High-End Computing (HEC) Program through the NASA Advanced Supercomputing (NAS) Division at Ames Research Center. We acknowledge PRACE for awarding us access to Curie at GENCI@CEA, France. Storage and computing resources at the University of Copenhagen HPC centre, funded in part by Villum Fonden (VKR023406), were used to carry out part of the data analysis.

\section*{Data Availability}

The $145$ spectra of the sources studied in the paper, and the parameters present in Appendix~\ref{app_table_clumps}, can be obtained from a dedicated public URL (\url{https://www.erda.dk/vgrid/larson-relations/}).

\bibliographystyle{mnras}
\bibliography{larson_relation}

\begin{appendix}

\section{$\rm N_{2}H^{+}(1-0)$ Spectra} \label{app_n2hp_sprectra}

\begin{figure*}
\centering
\includegraphics[width=0.98\textwidth]{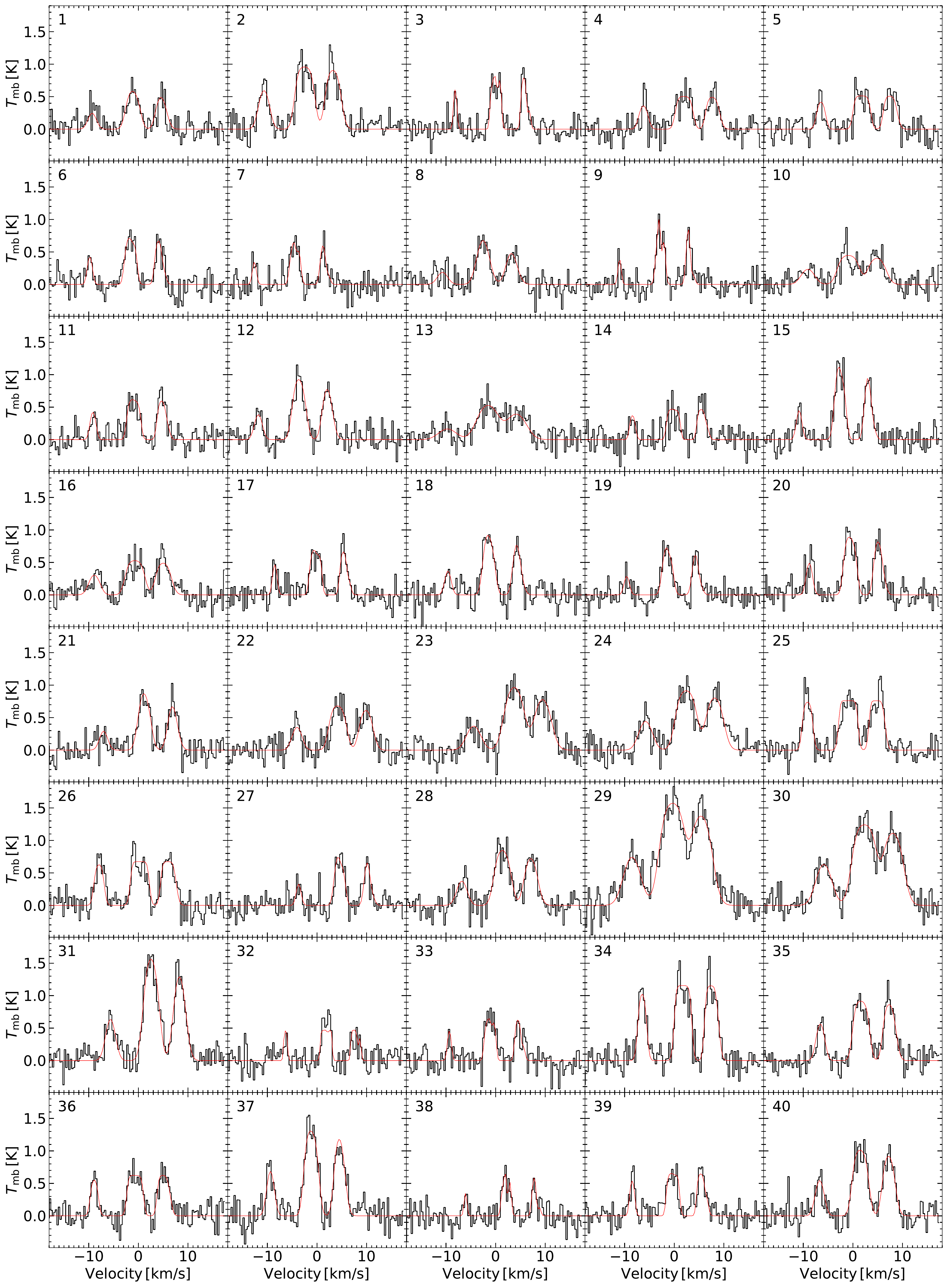}
\caption{$\rm N_{2}H^{+} (1-0)$ spectra of the 145 massive clumps. The black and red lines are the spectra and PySpecKit Gaussian fits, respectively. }
\label{fig_app_spec}
\end{figure*}

\begin{figure*}
\centering
\includegraphics[width=0.98\textwidth]{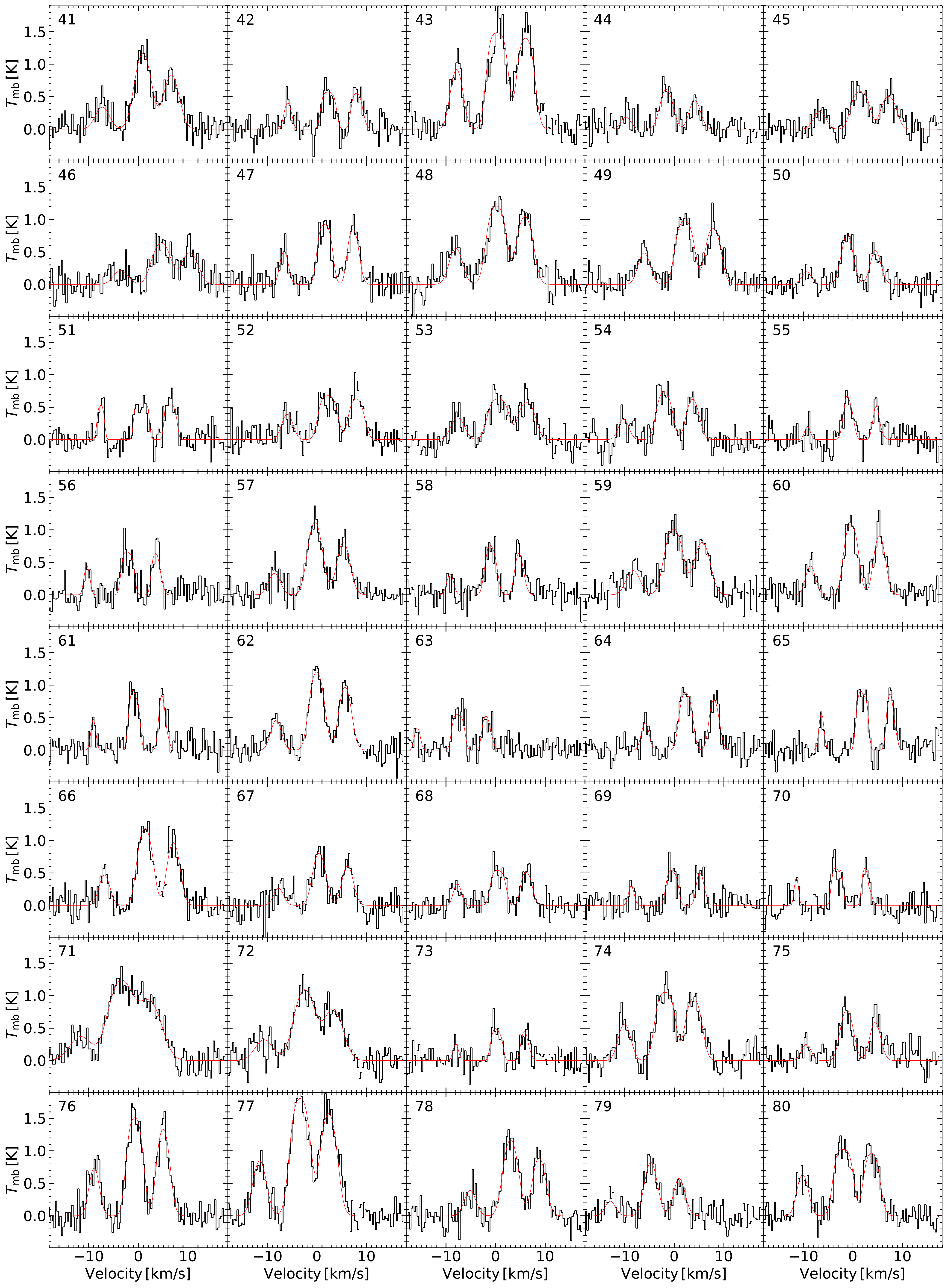}
\caption{$\rm N_{2}H^{+} (1-0)$ spectra. Figure \ref{fig_app_spec} --- Continued.}
\end{figure*}

\begin{figure*}
\centering
\includegraphics[width=0.98\textwidth]{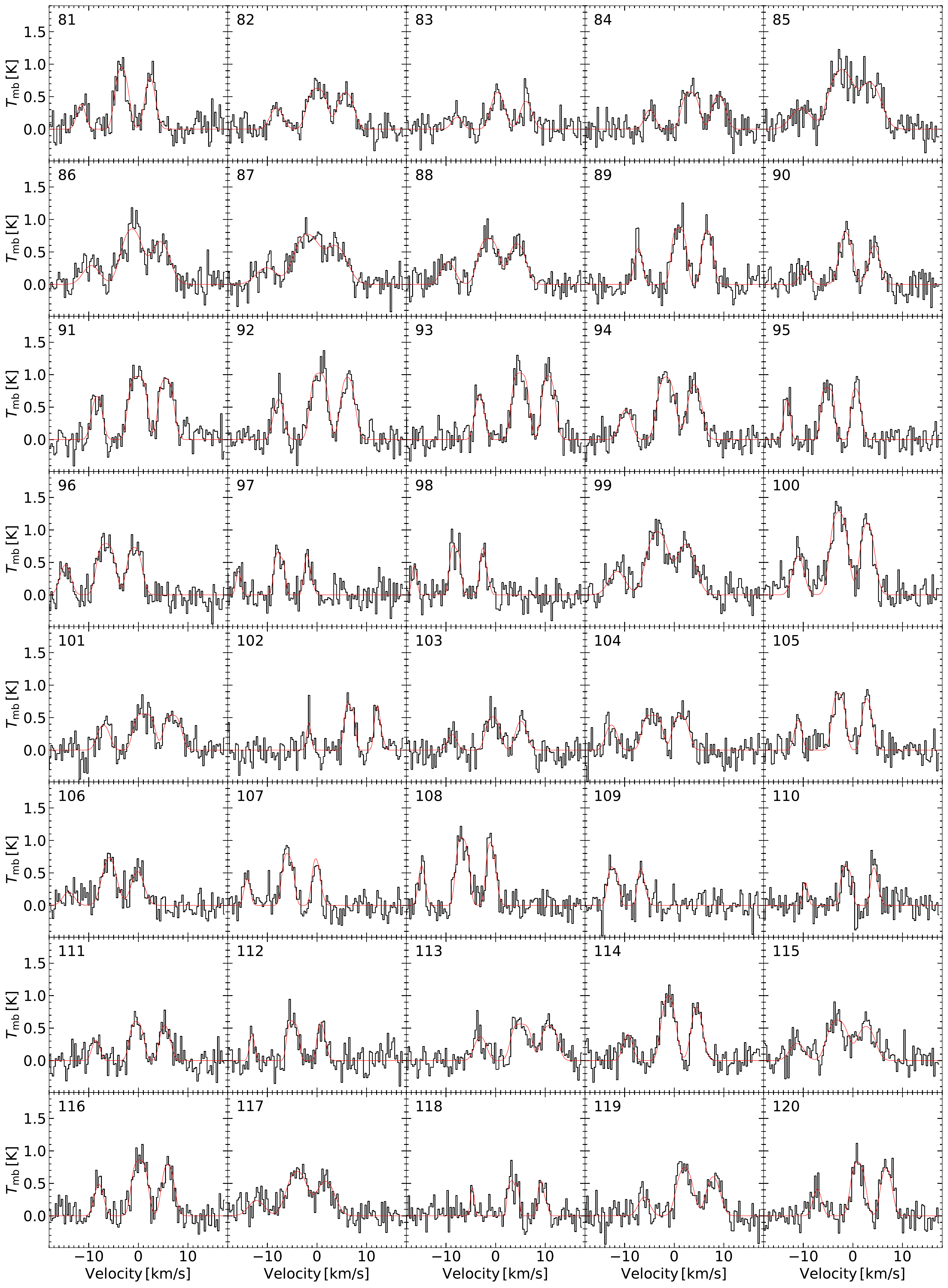}
\caption{$\rm N_{2}H^{+}(1-0)$ spectra. Figure \ref{fig_app_spec} --- Continued.}
\end{figure*}

\begin{figure*}
\centering
\includegraphics[width=0.98\textwidth]{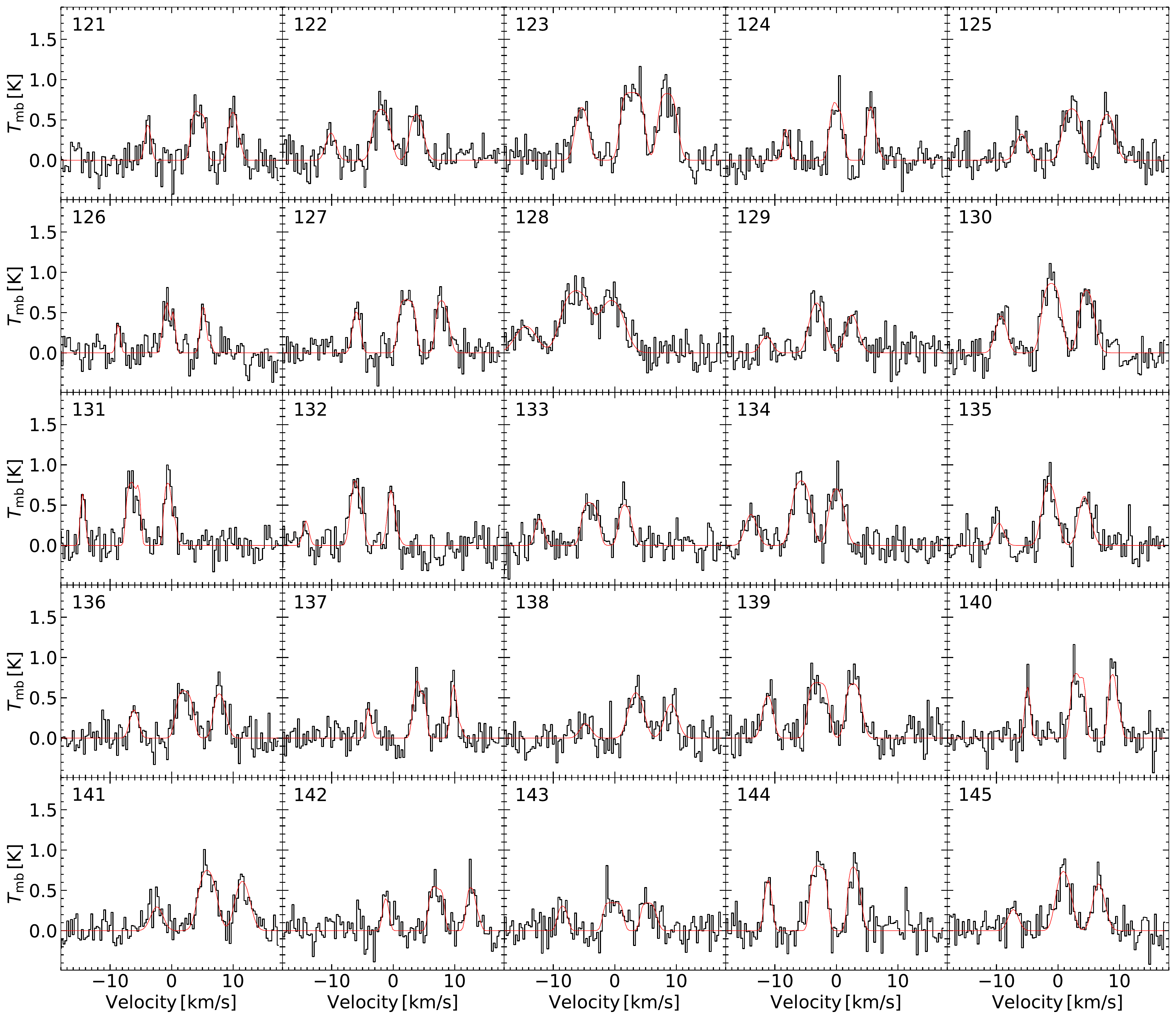}
\caption{$\rm N_{2}H^{+} (1-0)$ spectra. Figure \ref{fig_app_spec} --- Continued.}
\end{figure*}

\clearpage

\section{Clump Properties} \label{app_clump_properties}

The table of the properties of the 145 clumps studied in the paper.

\begin{table*}
\centering
\caption{The properties of the 145 clumps studied in the paper. An electronic version of this table is available at \url{https://www.erda.dk/vgrid/larson-relations/}.}
\begin{tabular}{ccccccccc}
\hline
\hline
$\rm Clump$ & $ R $ & $ M_{\rm SED} $ & $ T $ & $\sigma_{\rm N_{2}H^{+}}$ & $\alpha_{\rm vir,N_{2}H^{+}}$ & $ M_{\rm main} $ & $\sigma_{\rm main}$ & $\alpha_{\rm vir,main}$ \\
& $\rm [pc]$ & $\rm [M_{\odot}]$ & $\rm [K]$ & $\rm [km \, \rm s^{-1}]$ & & $\rm [M_{\odot}]$ & $\rm [km \, \rm s^{-1}]$ & \\
\hline
1 & 0.095 & 16.9 & 24.5 & 0.785 & 4.01 & 8.99 & 0.579 & 1.24 \\
2 & 0.111 & 88.0 & 19.6 & 0.932 & 1.27 & 19.4 & 0.807 & 2.4 \\
3 & 0.0882 & 11.1 & 16.0 & 0.279 & 0.72 & 10.3 & 0.416 & 0.912 \\
4 & 0.0339 & 1.2 & 16.2 & 0.714 & 16.8 & 5.42 & 0.429 & 0.921 \\
5 & 0.0254 & 0.509 & 24.2 & 0.64 & 23.8 & 3.1 & 0.377 & 1.4 \\
6 & 0.0482 & 3.78 & 14.1 & 0.452 & 3.03 & 6.65 & 0.641 & 1.1 \\
7 & 0.0504 & 2.15 & 19.5 & 0.372 & 3.77 & 3.55 & 0.535 & 2.09 \\
8 & 0.158 & 78.2 & 22.5 & 1.0 & 2.35 & 29.7 & 1.01 & 3.02 \\
9 & 0.127 & 23.1 & 16.4 & 0.321 & 0.661 & 21.3 & 0.594 & 0.894 \\
10 & 0.0413 & 1.68 & 29.3 & 1.19 & 40.1 & 8.92 & 1.43 & 12.4 \\
11 & 0.0502 & 1.95 & 17.7 & 0.499 & 7.48 & 5.61 & 0.497 & 1.59 \\
12 & 0.0733 & 9.28 & 22.1 & 0.746 & 5.12 & 11.6 & 1.02 & 4.27 \\
13 & 0.226 & 343 & 20.0 & 1.7 & 2.22 & 48.0 & 1.59 & 14.6 \\
14 & 0.043 & 4.3 & 17.0 & 0.43 & 2.15 & 3.12 & 0.424 & 2.43 \\
15 & 0.102 & 23.7 & 15.8 & 0.534 & 1.43 & 15.6 & 0.766 & 2.48 \\
16 & 0.031 & 22.4 & 13.3 & 0.957 & 1.47 & 8.92 & 1.43 & 9.29 \\
17 & 0.0734 & 10.3 & 14.8 & 0.393 & 1.28 & 9.77 & 0.653 & 2.32 \\
18 & 0.069 & 3.17 & 32.5 & 0.608 & 9.37 & 10.2 & 0.853 & 4.82 \\
19 & 0.177 & 66.0 & 22.1 & 0.513 & 0.822 & 28.3 & 1.03 & 3.77 \\
20 & 0.0848 & 15.5 & 13.5 & 0.568 & 2.05 & 12.8 & 0.464 & 0.857 \\
21 & 0.107 & 31.9 & 24.5 & 0.791 & 2.43 & 14.5 & 0.876 & 3.75 \\
22 & 0.0967 & 9.35 & 25.7 & 0.893 & 9.58 & 10.5 & 0.665 & 5.62 \\
23 & 0.0933 & 47.0 & 20.8 & 1.33 & 4.11 & 20.7 & 1.53 & 11.1 \\
24 & 0.0914 & 77.5 & 18.8 & 1.22 & 2.04 & 14.7 & 1.2 & 8.67 \\
25 & 0.0766 & 22.0 & 14.4 & 0.609 & 1.5 & 17.9 & 0.625 & 0.747 \\
26 & 0.0885 & 11.6 & 20.9 & 0.627 & 3.49 & 9.51 & 0.959 & 8.93 \\
27 & 0.059 & 2.06 & 21.4 & 0.468 & 7.28 & 3.49 & 0.66 & 9.13 \\
28 & 0.0676 & 3.89 & 26.0 & 0.964 & 18.8 & 12.3 & 1.12 & 5.0 \\
29 & 0.0976 & 95.1 & 21.3 & 1.56 & 2.91 & 32.2 & 2.12 & 10.0 \\
30 & 0.111 & 97.0 & 20.0 & 1.42 & 2.67 & 19.8 & 1.55 & 6.39 \\
31 & 0.117 & 65.6 & 14.7 & 0.922 & 1.76 & 24.0 & 1.19 & 3.21 \\
32 & 0.0956 & 12.9 & 12.8 & 0.228 & 0.448 & 13.6 & 0.438 & 0.602 \\
33 & 0.0495 & 3.18 & 26.2 & 0.332 & 1.99 & 3.38 & 0.532 & 4.44 \\
34 & 0.0999 & 87.2 & 19.7 & 0.586 & 0.458 & 14.1 & 1.18 & 5.71 \\
35 & 0.0579 & 11.1 & 31.8 & 0.737 & 3.28 & 2.67 & 0.638 & 2.57 \\
36 & 0.082 & 11.2 & 14.1 & 0.539 & 2.48 & 10.5 & 0.502 & 1.26 \\
37 & 0.0691 & 14.4 & 21.6 & 0.691 & 2.66 & 21.5 & 1.04 & 3.55 \\
38 & 0.0733 & 35.3 & 8.88 & 0.269 & 0.174 & 5.77 & 0.409 & 0.896 \\
39 & 0.0991 & 121 & 9.02 & 0.383 & 0.139 & 8.47 & 0.416 & 1.49 \\
40 & 0.0579 & 11.1 & 31.9 & 0.796 & 3.85 & 2.67 & 0.638 & 2.57 \\
41 & 0.134 & 174 & 13.5 & 1.36 & 1.66 & 28.4 & 1.4 & 5.84 \\
42 & 0.122 & 6.16 & 19.3 & 0.61 & 8.61 & 18.4 & 0.778 & 2.29 \\
43 & 0.0948 & 43.8 & 21.5 & 1.09 & 2.99 & 35.2 & 1.83 & 4.86 \\
44 & 0.0939 & 20.6 & 23.8 & 0.959 & 4.89 & 13.4 & 1.01 & 4.82 \\
45 & 0.204 & 122 & 22.1 & 1.12 & 2.41 & 59.6 & 1.06 & 3.6 \\
46 & 0.185 & 81.1 & 22.8 & 1.39 & 5.16 & 48.6 & 1.52 & 7.81 \\
47 & 0.224 & 304 & 19.8 & 0.746 & 0.476 & 49.8 & 0.939 & 3.99 \\
48 & 0.22 & 207 & 21.9 & 1.21 & 1.82 & 105 & 2.22 & 7.03 \\
49 & 0.163 & 99.7 & 22.8 & 1.09 & 2.26 & 43.3 & 1.4 & 2.92 \\
50 & 0.183 & 74.8 & 16.1 & 0.956 & 2.59 & 46.4 & 0.997 & 3.68 \\
51 & 0.127 & 8.11 & 25.5 & 0.425 & 3.29 & 16.3 & 0.764 & 3.19 \\
52 & 0.37 & 345 & 18.9 & 0.986 & 1.21 & 182 & 1.7 & 3.02 \\
53 & 0.251 & 153 & 24.7 & 1.19 & 2.7 & 89.9 & 2.21 & 9.46 \\
54 & 0.0592 & 3.32 & 24.7 & 0.962 & 19.2 & 8.24 & 0.834 & 5.83 \\
55 & 0.0295 & 0.568 & 21.5 & 0.556 & 18.7 & 2.16 & 0.368 & 0.891 \\
56 & 0.0438 & 5.76 & 17.9 & 0.485 & 2.07 & 4.46 & 0.52 & 2.33 \\
57 & 0.0323 & 3.38 & 24.0 & 1.15 & 14.6 & 4.85 & 1.22 & 11.2 \\
58 & 0.0765 & 16.7 & 23.7 & 0.638 & 2.16 & 4.8 & 1.01 & 13.1 \\
59 & 0.119 & 60.3 & 19.8 & 1.3 & 3.88 & 15.7 & 1.64 & 15.1 \\
60 & 0.0594 & 23.3 & 18.6 & 0.904 & 2.42 & 11.7 & 0.89 & 5.76 \\
\hline
\hline
\end{tabular}
\label{app_table_clumps}
\end{table*}

\begin{table*}
\centering
\caption{Table \ref{app_table_clumps} -- continued. }
\begin{tabular}{ccccccccc}
\hline
\hline
$\rm Clump$ & $ R $ & $ M_{\rm SED} $ & $ T $ & $\sigma_{\rm N_{2}H^{+}}$ & $\alpha_{\rm vir,N_{2}H^{+}}$ & $ M_{\rm main} $ & $\sigma_{\rm main}$ & $\alpha_{\rm vir,main}$ \\
& $\rm [pc]$ & $\rm [M_{\odot}]$ & $\rm [K]$ & $\rm [km \, \rm s^{-1}]$ & & $\rm [M_{\odot}]$ & $\rm [km \, \rm s^{-1}]$ & \\
\hline
61 & 0.0502 & 4.11 & 30.1 & 0.445 & 2.82 & 5.54 & 0.75 & 4.95 \\
62 & 0.046 & 6.19 & 21.6 & 0.999 & 8.63 & 13.7 & 1.26 & 3.68 \\
63 & 0.0578 & 6.08 & 21.2 & 0.5 & 2.76 & 3.05 & 0.993 & 14.5 \\
64 & 0.0552 & 3.32 & 19.4 & 0.71 & 9.75 & 6.6 & 0.511 & 1.68 \\
65 & 0.0991 & 16.1 & 19.0 & 0.429 & 1.31 & 11.3 & 1.02 & 2.46 \\
66 & 0.088 & 25.9 & 22.9 & 1.01 & 4.06 & 12.9 & 1.69 & 20.2 \\
67 & 0.102 & 30.2 & 24.1 & 0.951 & 3.57 & 13.2 & 0.855 & 1.06 \\
68 & 0.102 & 13.2 & 18.5 & 0.682 & 4.16 & 11.1 & 0.335 & 1.09 \\
69 & 0.086 & 37.0 & 13.7 & 0.51 & 0.702 & 5.89 & 0.576 & 3.88 \\
70 & 0.0504 & 1.73 & 30.1 & 0.347 & 4.09 & 4.2 & 0.475 & 3.35 \\
71 & 0.0954 & 45.4 & 25.3 & 2.12 & 10.9 & 37.7 & 2.0 & 8.76 \\
72 & 0.093 & 53.1 & 19.0 & 1.74 & 6.17 & 27.5 & 1.99 & 7.77 \\
73 & 0.0789 & 2.1 & 19.5 & 0.488 & 10.4 & 10.0 & 0.448 & 0.566 \\
74 & 0.087 & 72.2 & 13.0 & 1.12 & 1.76 & 10.6 & 1.09 & 9.78 \\
75 & 0.127 & 75.3 & 14.0 & 0.932 & 1.71 & 16.3 & 0.712 & 2.7 \\
76 & 0.0663 & 283 & 10.3 & 0.935 & 0.238 & 11.5 & 1.1 & 5.64 \\
77 & 0.0733 & 22.1 & 20.6 & 1.23 & 5.78 & 50.5 & 1.22 & 1.22 \\
78 & 0.0995 & 34.5 & 21.2 & 1.1 & 4.07 & 14.3 & 1.43 & 10.6 \\
79 & 0.0933 & 22.2 & 21.6 & 1.01 & 4.96 & 10.4 & 0.996 & 3.87 \\
80 & 0.087 & 58.2 & 13.5 & 0.995 & 1.72 & 10.6 & 1.09 & 9.78 \\
81 & 0.0972 & 95.1 & 13.9 & 0.815 & 0.789 & 11.7 & 1.12 & 6.5 \\
82 & 0.273 & 314 & 16.9 & 1.14 & 1.31 & 81.9 & 1.47 & 4.31 \\
83 & 0.186 & 45.0 & 26.6 & 0.978 & 4.6 & 41.2 & 1.13 & 1.29 \\
84 & 0.198 & 75.9 & 25.4 & 0.945 & 2.71 & 42.1 & 1.3 & 4.64 \\
85 & 0.195 & 218 & 20.3 & 1.84 & 3.52 & 56.1 & 2.21 & 10.2 \\
86 & 0.397 & 582 & 17.7 & 1.58 & 1.99 & 205 & 1.74 & 2.71 \\
87 & 0.364 & 189 & 26.2 & 1.96 & 8.56 & 216 & 1.85 & 4.85 \\
88 & 0.729 & 5220 & 12.5 & 1.25 & 0.252 & 743 & 1.44 & 1.54 \\
89 & 0.0628 & 9.02 & 19.9 & 0.677 & 3.72 & 9.52 & 0.623 & 1.21 \\
90 & 0.0933 & 46.3 & 18.4 & 0.945 & 2.09 & 12.8 & 1.41 & 11.2 \\
91 & 0.0896 & 21.4 & 17.5 & 0.877 & 3.75 & 13.3 & 1.17 & 10.5 \\
92 & 0.1 & 54.0 & 23.8 & 0.891 & 1.72 & 35.5 & 1.43 & 3.88 \\
93 & 0.108 & 109 & 13.3 & 0.788 & 0.714 & 38.8 & 0.785 & 1.25 \\
94 & 0.0803 & 13.1 & 23.0 & 1.04 & 7.7 & 18.8 & 1.11 & 4.97 \\
95 & 0.0641 & 8.04 & 17.1 & 0.506 & 2.38 & 11.7 & 0.652 & 1.4 \\
96 & 0.0625 & 4.54 & 30.9 & 0.957 & 14.7 & 22.9 & 1.22 & 2.69 \\
97 & 0.0891 & 12.1 & 13.5 & 0.41 & 1.44 & 7.88 & 0.473 & 1.42 \\
98 & 0.0777 & 3.52 & 16.4 & 0.43 & 4.76 & 6.43 & 0.454 & 2.81 \\
99 & 0.0979 & 34.3 & 22.5 & 1.57 & 8.14 & 19.0 & 1.63 & 14.0 \\
100 & 0.0563 & 13.0 & 22.0 & 1.04 & 5.43 & 6.94 & 1.16 & 7.32 \\
101 & 0.0881 & 13.6 & 14.6 & 1.01 & 7.65 & 8.56 & 0.757 & 2.13 \\
102 & 0.0671 & 7.18 & 12.5 & 0.398 & 1.72 & 11.7 & 0.456 & 0.704 \\
103 & 0.0581 & 12.8 & 19.9 & 0.804 & 3.41 & 3.12 & 0.686 & 9.34 \\
104 & 0.0637 & 13.0 & 16.9 & 0.932 & 4.96 & 10.2 & 1.06 & 7.54 \\
105 & 0.105 & 21.4 & 15.5 & 0.669 & 2.56 & 13.9 & 0.746 & 2.33 \\
106 & 0.0829 & 15.6 & 28.1 & 1.13 & 7.9 & 6.86 & 1.42 & 8.15 \\
107 & 0.0939 & 60.9 & 19.6 & 0.553 & 0.548 & 7.18 & 0.952 & 3.82 \\
108 & 0.13 & 269 & 12.0 & 0.592 & 0.197 & 20.5 & 1.17 & 4.69 \\
109 & 0.0725 & 94.5 & 10.2 & 0.439 & 0.172 & 5.77 & 0.387 & 1.47 \\
110 & 0.114 & 20.4 & 14.0 & 0.446 & 1.29 & 15.1 & 0.731 & 1.74 \\
111 & 0.0299 & 0.613 & 37.7 & 0.705 & 28.2 & 1.25 & 0.493 & 5.4 \\
112 & 0.098 & 2.51 & 25.4 & 0.456 & 9.45 & 11.3 & 0.482 & 0.882 \\
113 & 0.0972 & 20.9 & 19.8 & 1.0 & 5.45 & 12.8 & 0.704 & 2.5 \\
114 & 0.102 & 48.4 & 18.7 & 0.872 & 1.87 & 13.3 & 1.15 & 9.7 \\
115 & 0.192 & 164 & 18.0 & 1.35 & 2.46 & 40.0 & 1.49 & 3.86 \\
116 & 0.135 & 61.2 & 14.3 & 0.811 & 1.69 & 23.2 & 1.33 & 9.66 \\
117 & 0.544 & 1160 & 18.6 & 1.52 & 1.27 & 519 & 1.95 & 2.41 \\
118 & 0.0647 & 6.95 & 19.1 & 0.319 & 1.1 & 5.88 & 0.469 & 2.21 \\
119 & 0.0775 & 26.7 & 17.4 & 1.08 & 3.94 & 10.7 & 1.36 & 16.0 \\
120 & 0.0362 & 6.68 & 17.9 & 0.781 & 3.85 & 8.24 & 0.684 & 2.58 \\
\hline
\hline
\end{tabular}
\end{table*}

\begin{table*}
\centering
\caption{Table \ref{app_table_clumps} -- continued. }
\begin{tabular}{ccccccccc}
\hline
\hline
$\rm Clump$ & $ R $ & $ M_{\rm SED} $ & $ T $ & $\sigma_{\rm N_{2}H^{+}}$ & $\alpha_{\rm vir,N_{2}H^{+}}$ & $ M_{\rm main} $ & $\sigma_{\rm main}$ & $\alpha_{\rm vir,main}$ \\
& $\rm [pc]$ & $\rm [M_{\odot}]$ & $\rm [K]$ & $\rm [km \, \rm s^{-1}]$ & & $\rm [M_{\odot}]$ & $\rm [km \, \rm s^{-1}]$ & \\
\hline
121 & 0.0848 & 5.26 & 15.0 & 0.409 & 3.14 & 9.92 & 0.69 & 3.56 \\
122 & 0.0894 & 10.9 & 14.1 & 0.722 & 4.97 & 12.4 & 0.578 & 2.27 \\
123 & 0.0837 & 7.16 & 31.1 & 0.81 & 8.93 & 19.6 & 0.795 & 4.38 \\
124 & 0.0972 & 38.8 & 12.9 & 0.447 & 0.582 & 7.91 & 0.519 & 1.44 \\
125 & 0.0512 & 1.23 & 29.1 & 0.823 & 32.6 & 7.29 & 0.701 & 1.98 \\
126 & 0.0612 & 21.7 & 8.55 & 0.301 & 0.297 & 5.59 & 0.338 & 0.649 \\
127 & 0.0229 & 2.97 & 17.9 & 0.507 & 2.31 & 1.94 & 0.396 & 2.08 \\
128 & 0.0944 & 15.6 & 21.3 & 1.58 & 17.5 & 19.6 & 1.33 & 10.2 \\
129 & 0.0458 & 3.5 & 20.7 & 0.859 & 11.2 & 4.4 & 0.973 & 9.45 \\
130 & 0.0314 & 0.937 & 29.9 & 0.811 & 25.6 & 8.36 & 0.829 & 2.18 \\
131 & 0.0822 & 89.0 & 12.7 & 0.309 & 0.102 & 13.0 & 0.493 & 1.25 \\
132 & 0.0656 & 4.41 & 19.6 & 0.475 & 3.9 & 6.33 & 0.499 & 2.11 \\
133 & 0.0939 & 61.9 & 8.2 & 0.59 & 0.613 & 13.4 & 0.751 & 2.27 \\
134 & 0.0131 & 4.51 & 16.3 & 0.862 & 2.5 & 4.42 & 0.548 & 0.793 \\
135 & 0.0651 & 2.93 & 21.3 & 0.769 & 15.2 & 8.78 & 0.742 & 2.55 \\
136 & 0.0931 & 22.4 & 19.5 & 0.612 & 1.81 & 9.28 & 0.39 & 0.436 \\
137 & 0.0244 & 0.554 & 31.6 & 0.368 & 6.95 & 0.924 & 0.577 & 10.1 \\
138 & 0.0554 & 2.17 & 25.5 & 0.871 & 22.5 & 1.68 & 0.285 & 2.57 \\
139 & 0.0513 & 12.9 & 14.6 & 0.685 & 2.16 & 9.98 & 0.618 & 2.68 \\
140 & 0.0893 & 23.5 & 19.9 & 0.354 & 0.556 & 13.2 & 0.593 & 1.47 \\
141 & 0.054 & 2.55 & 37.3 & 0.939 & 21.7 & 5.6 & 1.26 & 19.3 \\
142 & 0.0543 & 5.04 & 19.7 & 0.411 & 2.11 & 4.3 & 0.459 & 2.51 \\
143 & 0.0952 & 39.9 & 8.55 & 0.552 & 0.847 & 12.0 & 0.666 & 0.651 \\
144 & 0.015 & 0.696 & 27.1 & 0.484 & 5.87 & 1.73 & 0.478 & 2.71 \\
145 & 0.0311 & 0.436 & 24.3 & 0.794 & 52.2 & 5.9 & 0.603 & 1.43 \\
\hline
\hline
\end{tabular}
\end{table*}

\clearpage

\twocolumn

\section{The Effect of Band Merging on the Mass Estimate} \label{app_band_merging}

\begin{figure}
\centering
\includegraphics[width=0.48\textwidth]{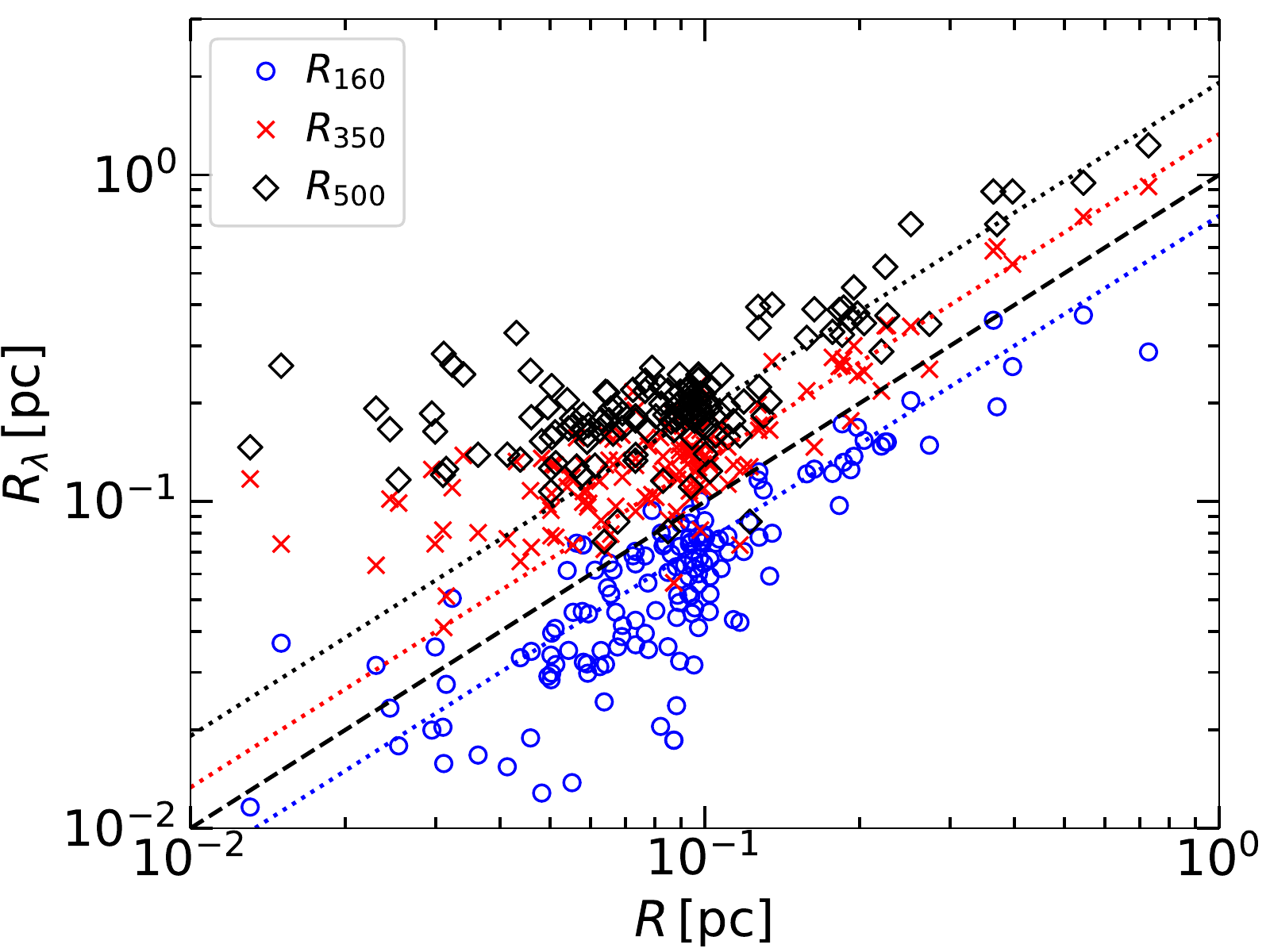}
\caption{Deconvolved source sizes at different wavelengths, $R_{\lambda}$, as a function of the source radius, $R$, that is the deconvolved size at 250~$\mu$m. The black dashed line shows $R_{\lambda}=R$, so it corresponds to the case of $\lambda=250$~$\mu$m, while the three dotted lines show the predicted source sizes at the other wavelengths if they were equal to the size at 250~$\mu$m rescaled by the ratio of the beam sizes.}
\label{fig_radius_comparison}
\end{figure}

When fitting the SED of the dust emission under the assumption of uniform temperature, source masses are underestimated if significant temperature variations exist within the source or in projection on the line of sight, because the dust emission is weighted towards the warmer regions and underestimates the amount of colder dust. Dust emissivity variations may also exist, which can lead to an underestimate or an overestimate of the source mass depending on the dust properties within the source. However, the masses may be overestimated when projection effects in the line of sight of the sources are strong, such that the photometric background subtraction cannot properly disentangle the actual 3D clump. In this study, projection effects may be important, as the mass of the main 3D clump is on average only one third of the total mass in the line of sight. 

Uncertainties in the mass estimation may also be induced by the method of computing the SEDs, when the fluxes of different wavelengths are evaluated from different areas and resolutions, as in the case of Hi-GAL's band merging procedure \citep{Elia+2017}, which we have followed in our synthetic catalog \citep{Lu+2021}. In Hi-GAL, detection and photometry, including background subtraction, are done separately at different bands at different resolutions. The detections from different bands are then assumed to be the same source if the center of the shorter-wavelength detection is within the radius of the longer-wavelength counterpart \citep[see][for further details]{Lu+2021}. Thus, the SED fluxes at different bands come from sources that generally have different sizes and may only partially overlap with each other. In addition, although multiple shorter-wavelength sources may be found within the longer-wavelength ones, only one source per wavelength (the closest one to the center of the longer-wavelength counterpart) is retained into the SED. Because of the size differences, the fluxes at 350 and 500~$\mu$m are rescaled (reduced) by the ratio of their deconvolved sizes with that of the 250~$\mu$m counterpart. On the other hand, the flux of the 160~$\mu$m counterparts is not rescaled (increased) even if their deconvolved size is generally smaller. 

This band-merging procedure introduces two main uncertainties in the source mass estimation: 
\begin{enumerate}
\item the shorter-wavelength flux is generally underestimated because only one shorter-wavelength detection is retained even when multiple ones are found within the area of the longer-wavelength counterpart. It is further underestimated at 160~$\mu$m because it is not rescaled. As a result, the temperature is underestimated and the mass overestimated.   
\item in cases with a large difference between the deconvolved sizes at different wavelengths, the short wavelength flux arises from a much smaller hot region within the larger and colder longer-wavelength counterpart. When the longer-wavelength flux is rescaled to that smaller size, it underestimates the actual flux from that hot region, leading to an estimated temperature that is even higher than that of the hot region, and thus an underestimated mass.
\end{enumerate}

At large source sizes, $R$, Figure~\ref{fig_mass_comparison} shows that the estimated source mass, $M_{\rm SED}$, is on average larger than the mass of the associated 3D clump, $M_{\rm main}$, thus the first effect of the band-merging procedure and the projection effects must be dominant compared to the second effect of the band-merging or the effect of the uniform temperature approximation. We find that $M_{\rm SED}$ is on average of the same order as the total mass in the line of sight, suggesting that projection effects are the main factor. On the contrary, at small sizes, $M_{\rm SED}$ is generally smaller than $M_{\rm main}$, so the second effect must be the dominant one, meaning that the deconvolved sizes at different wavelengths must be very different.   
This is confirmed in Figure~\ref{fig_radius_comparison}, where we show the deconvolved sizes at different wavelengths versus that at 250~$\mu$m, which is our source size, $R$. While at large $R$ the deconvolved sizes at different wavelengths scale linearly with R, there is a clear transition below approximately 0.1~pc, where the deconvolved sizes at 350 and 500~$\mu$m stop decreasing with decreasing $R$ (they remain of the order of their respective beams). As a further confirmation of this effect, we find that the brightness temperature of the sources increases systematically with increasing $M_{\rm main}/M_{\rm SED}$ and decreasing $R$.

\end{appendix}

\bsp	
\label{lastpage}
\end{document}